\documentclass[prl,twocolumn,showpacs,floatfix,amsmath,amssymb,superscriptaddress]{revtex4-1}
\usepackage{amsfonts}
\usepackage{stmaryrd}
\usepackage{bbm}
\usepackage{mathrsfs}
\usepackage{tipa}
\usepackage{amssymb}
\usepackage{txfonts}
\usepackage{graphicx}
\usepackage{dcolumn}
\usepackage{epstopdf}
\usepackage[colorlinks,linkcolor=blue,urlcolor=blue,citecolor=blue]{hyperref}
\usepackage{multirow}
\usepackage{subfigure}
\usepackage{url}

\begin{document}
\newcommand*{\cm}{cm$^{-1}$\,}
\newcommand*{\Tc}{T$_c$\,}

\title{Revealing a charge-density-wave gap in the predicted weak topological insulator HoSbTe}

\author{J. L. Liu}
\affiliation{Center for Advanced Quantum Studies and Department of Physics, Beijing Normal University, Beijing 100875, China}

\author{R. Liu}
\affiliation{Center for Advanced Quantum Studies and Department of Physics, Beijing Normal University, Beijing 100875, China}

\author{M. Yang}
\affiliation{Beijing National Laboratory for Condensed Matter Physics and Institute of Physics, Chinese Academy of Sciences, Beijing 100190, China}
\affiliation{Center of Materials Science and Optoelectronics Engineering, University of Chinese Academy of Sciences, Beijing 100049, China}
\affiliation{School of Physical Sciences, University of Chinese Academy of Sciences, Beijing 100190, China}

\author{L. Y. Cao}
\affiliation{Center for Advanced Quantum Studies and Department of Physics, Beijing Normal University, Beijing 100875, China}

\author{B. X. Gao}
\affiliation{Center for Advanced Quantum Studies and Department of Physics, Beijing Normal University, Beijing 100875, China}

\author{L. Wang}
\affiliation{Center for Advanced Quantum Studies and Department of Physics, Beijing Normal University, Beijing 100875, China}

\author{A. F. Fang}
\affiliation{Department of Physics, Beijing Normal University, Beijing 100875, China}

\author{Y. G. Shi}
\affiliation{Beijing National Laboratory for Condensed Matter Physics and Institute of Physics, Chinese Academy of Sciences, Beijing 100190, China}
\affiliation{Center of Materials Science and Optoelectronics Engineering, University of Chinese Academy of Sciences, Beijing  100049, China}
\affiliation{School of Physical Sciences, University of Chinese Academy of Sciences, Beijing 100190, China}

\author{Z. P. Yin}
\affiliation{Center for Advanced Quantum Studies and Department of Physics, Beijing Normal University, Beijing 100875, China}

\author{R. Y. Chen}
\affiliation{Center for Advanced Quantum Studies and Department of Physics, Beijing Normal University, Beijing 100875, China}

\begin{abstract}

HoSbTe was predicted to be a weak topological insulator, whose spin-orbit coupling (SOC) gaps are reported to be as large as hundreds of meV. Utilizing infrared spectroscopy, we find that the compound is of metallic nature from 350 K down to 10 K. Particularly, both of its itinerant carrier density and scattering rate are demonstrated to decrease with temperature cooling, which is responsible for the appearance of a broad hump feature in the temperature dependent resistivity around 200 K. More importantly, we reveal the appearance of a charge-density-wave (CDW) gap in addition to the SOC related gap. The energy scale of the CDW gap is identified to be 364 meV at 10 K, which shift to 252 meV at 350 K. The coexistence of CDW and SOC gaps in the same compound paves a new avenue to explore more intriguing physics.


\end{abstract}

\maketitle
\section{introduction}

Topological materials with nontrivial electronic bands have drawn tremendous interests in condensed matter physics due to their exotic physical properties and their potential applications\cite{RevModPhys.83.1057,RevModPhys.82.3045,RevModPhys.80.1083}. Huge progress has been made in the field and a large number of topological materials have been recognized. However topological materials with strong electronic correlation are much less explored and might host more interesting physics\cite{Pesin2010,Deng2017,Schoop2018,Peters2018,majorana,Wang2013}. 
The $WHM$ ($W$=Zr, Hf, or Lanthanides, $H$=Si, Ge, Sn, or Sb, and $M$=O, S, Se, or Te) family, initially noticed by theorist in the search for two-dimensional topological insulators\cite{Xu2015}, can serve as a good platform to study the interactions between topology, magnetism and other emergent instabilities.

Most of the $WHM$ compounds are reported to be Dirac nodal line semimetals\cite{DNL-ZrGeSe,DNL-ZrSiSe,Wang2021,DNL-ZrGeTe,DNL-ZrGeXc,DNL-ZrSiSeandZrSiTe,DNL-ZrSiS} or weak topological insulators\cite{Xu2015,TI-ZrSnTe}. However when $W$ are Lanthanides with $f$ electrons, magnetic orderings and Kondo effect could play a role in the system. For example, LaSbTe was suggested to be a topological insulator\cite{Singha2017} or nodal-line semimetal\cite{Wang2021} by different experiments.  Its counterpart CeSbTe with $f$ electrons from Ce is found to be a low-carrier-density Kondo semimetal with a Kondo temperature of 10 K\cite{Lv2019}, and its electronic structure could be severely modified by long range magnetic orders\cite{Schoop2018}. Meanwhile GdSbTe is demonstrated to be a nodal-line semimetal with a robust Dirac-like band structure across the AFM transition at 13 K\cite{Hosen2018}.

Among these $WHM$ materials HoSbTe was theoretically predicted to be a weak topological insulator\cite{Yang2020}, which is corroborated afterwards by angle-resolved photoemission spectroscopy (ARPES) experiments exhibiting energy gaps much larger than 200 meV along certain momentum directions\cite{Yue2020}. However the temperature dependent resistivity of HoSbTe exhibits a broad hump-like feature at around 200 K, indicating a bad-metal-like state at low temperatures, the underlying physics of which is yet to be revealed\cite{Yang2020}. The 4$f$ electrons of Ho atoms are believed to be generally localized and far away from the Fermi level, but an extremely large Sommerfeld coefficient $\gamma \sim$  382.2 mJ/mol$^{-1}$/K$^2$ is obtained by a specific heat measurement\cite{Yang2020}, which is much larger than that of CeSbTe ($\gamma$ = 41 mJ/mol$^{-1}$/K$^2$)\cite{Lv2019}, and comparable to many typical heavy fermion materials like CeCoIn5\cite{Petrovic_2001}. This seems to infer a very large Kondo hybridization strength and thus the itinerancy of the $f$ electrons in HoSbTe.

In order to resolve the existing enigma and further investigate the role played by 4$f$ electrons, we performed infrared spectroscopy measurements on single crystalline HoSbTe samples. Our results suggest that the hump feature in resistivity could be well explained by the competing effect of carrier density and scattering rate. In addition to the expected SOC gap we observe for the first time another temperature dependent gap whose value varies from 364 meV at 10K to 252 meV at 250K. We propose that this newly observed gap is caused by CDW ordering similar to those in CeSbTe\cite{Li2021}.

\section{experimental}

Single crystalline HoSbTe samples were synthesized by Sb-flux method\cite{Yang2020}.
The temperature dependent resistivity $\rho(T)$ and magnetization $\chi(T)$ up to 350 K were measured in a Quantum Design physical property measurement system.

Infrared spectroscopic studies were performed with a Bruker IFS 80 V in the frequency range from 30 to 50 000 \cm, on the as-grown shiny surfaces of HoSbTe which is the $ab$ plane of this quasi-two dimensional compound. In the measurement of frequency dependent reflectivity $R (\omega)$, either gold or aluminum coating techniques are adopted in order to eliminate the impact of microscopic surface textures of the single crystalline compound. The real part of the optical conductivity $\sigma_1(\omega)$  are derived through Kramers-Kronig transformation of the reflectivity R($\omega$), which is extrapolated by a Hagen-Rubens relation to zero at the low frequency end, and by $\omega^{-1.5}$ from 50 000 \cm to 800 000 \cm, and then by $\omega^{-4}$ for frequency higher than 800 000 \cm.

Theoretical calculations were carried out using the linearized-augmented plane-wave method implemented in WIEN2K\cite{S3}. The Perdew-Burke-Ernzerhof version of generalized-gradient-approximation (GGA) to the exchange correlation functional\cite{PhysRevLett.77.3865} was used. To move the $f$ orbitals of Ho atoms away from Fermi level, GGA+U method was adopted in our calculations, and we applied an effective Hubbard $U_{eff}=U-J_H$ = 7 eV to the Ho f orbitals. Because the AFM transition temperature of HoSbTe ($T_N$ $\sim$ 4 K) is very low, we ignore the magnetic order and force the calculations to non-magnetic solution. In self-consistent calculation we used 27$\times$27$\times$12 k-mesh grid, and in order to clearly show Fermi surfaces sections, 500$\times$500 k-mesh grid was used in each specified $kz$ plane. Spin-orbit coupling (SOC) was taken into account in our calculation.

\section{results and discussion}

\begin{figure}[htb]
\centering
  \includegraphics[width=5cm]{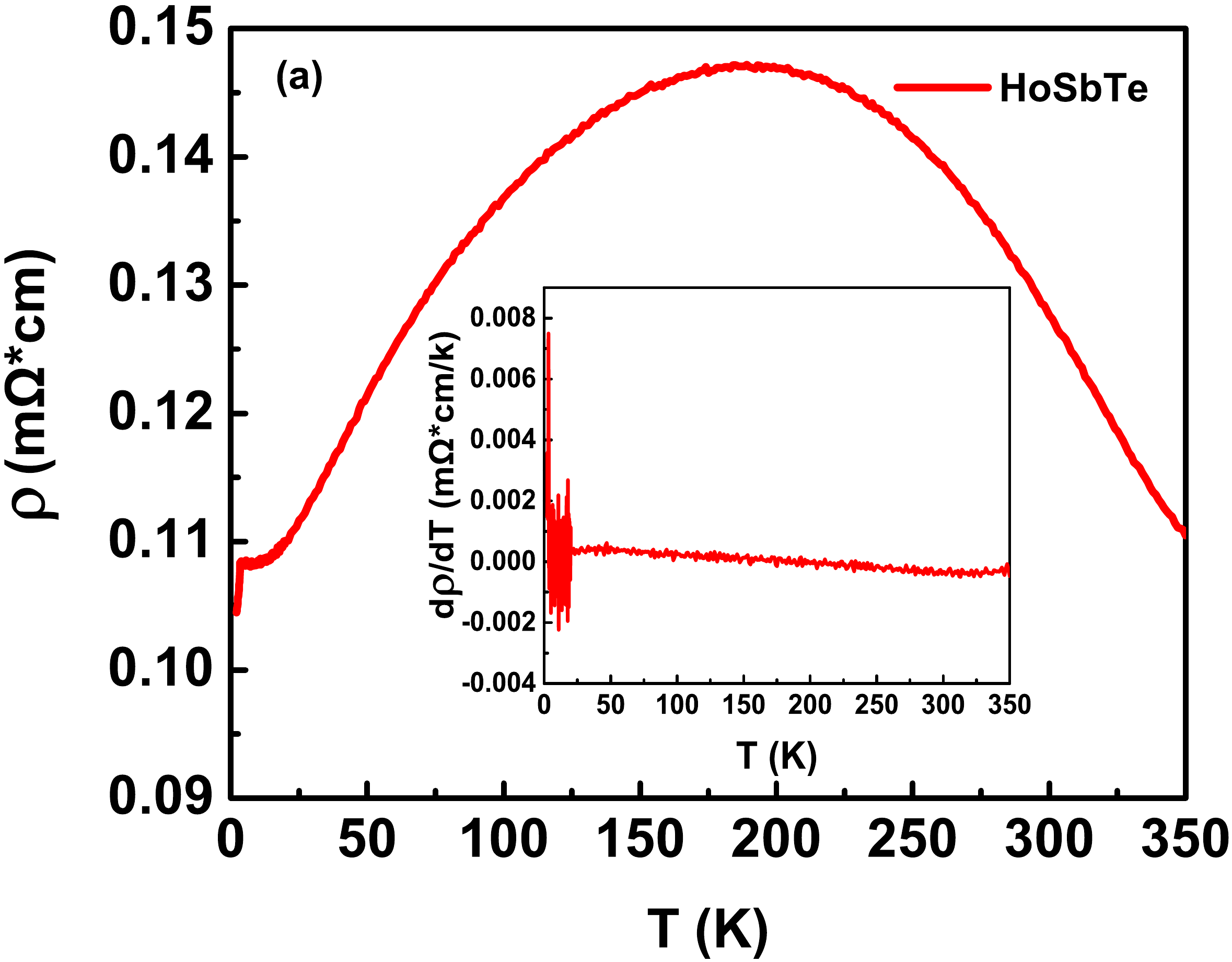}
  \includegraphics[width=4.9cm]{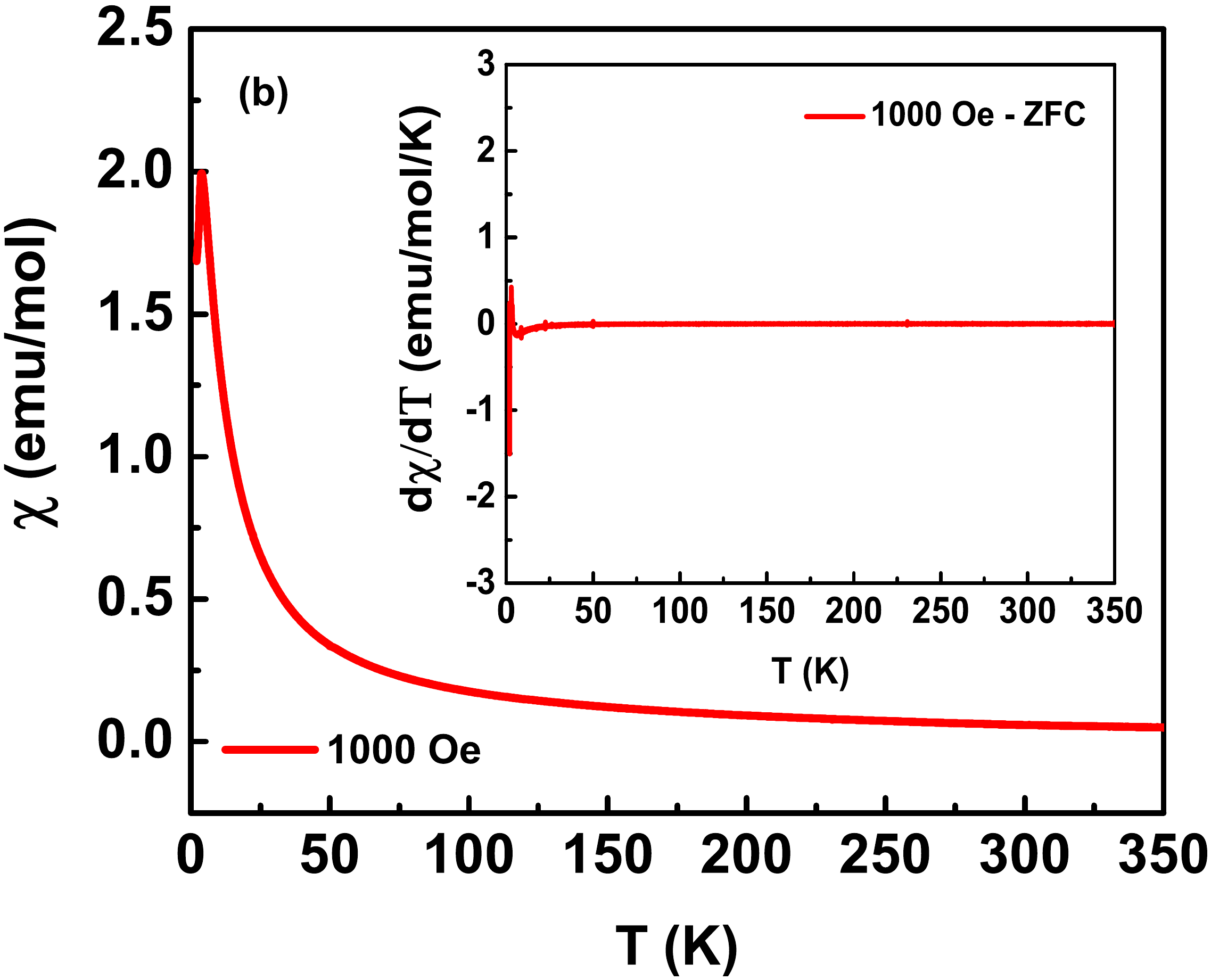}
  \caption{The in-plane (a) resistivity and (b) magnetic susceptibility as a function of temperature. The magnetic susceptibility is measured with H = 1000 Oe. The first order derivatives of resistivity and magnetic susceptibility are displayed in the inset of (a) and (b), respectively.}\label{Fig:1}
\end{figure}

Fig.~\ref{Fig:1}(a) and (b) display the temperature dependent resistivity $\rho(T)$ and magnetization $\chi(T)$, respectively. Although we have measured $\rho(T)$ and $\chi(T)$ up to a higher temperature, there is not much difference with the previous report\cite{Yang2020}: $\rho(T)$ shows a broad hump around 200 K and $\chi(T)$ follows the Curie-Wiess law down to at least 100 K; an antiferromagnetic phase transition occurs at around 4 K. In order to check if there are any other phase transitions that may lead to the ``insulator to metal'' change inferred by resistivity, the first-order derivatives $d\rho(T)/dT$ and $d\chi(T)/dT$ are calculated and displayed in the inset of Fig.~\ref{Fig:1}(a) and (b), respectively. The smooth derivative curves at relatively high temperatures obviously do not support existence of such transitions.

\begin{figure*}[htpb]
\centering

  \includegraphics[width=7cm]{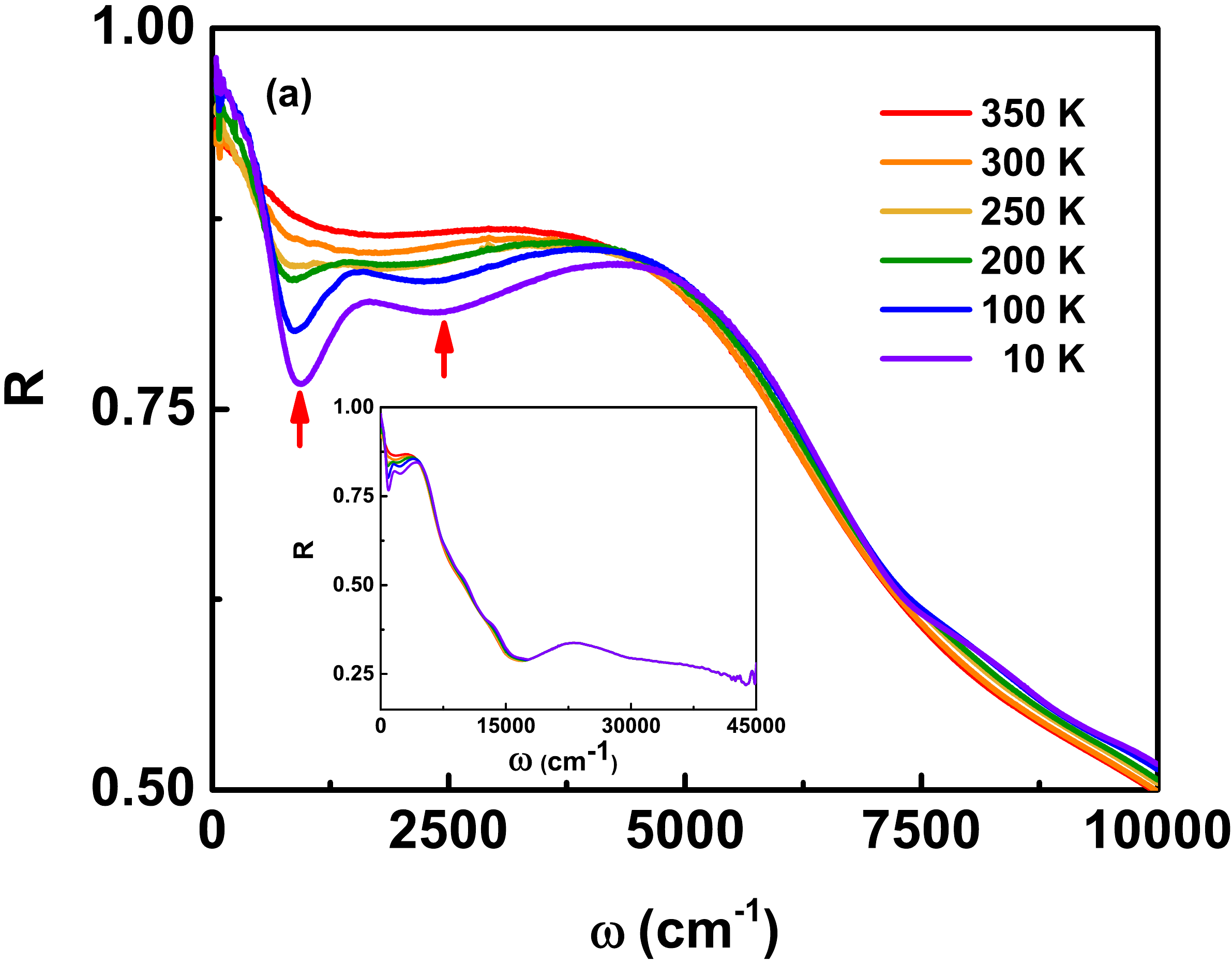}
  \includegraphics[width=7cm]{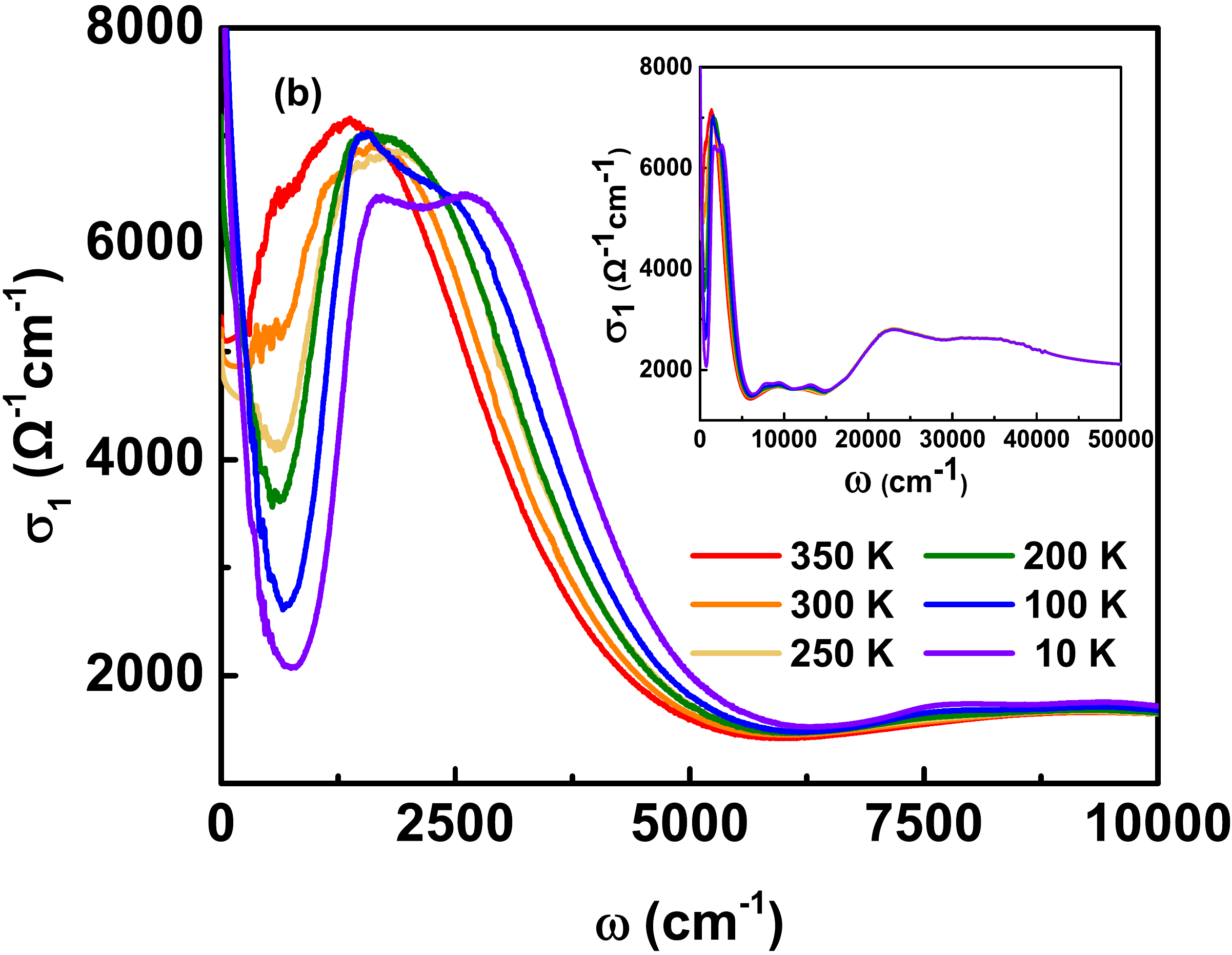}
  \includegraphics[width=7cm]{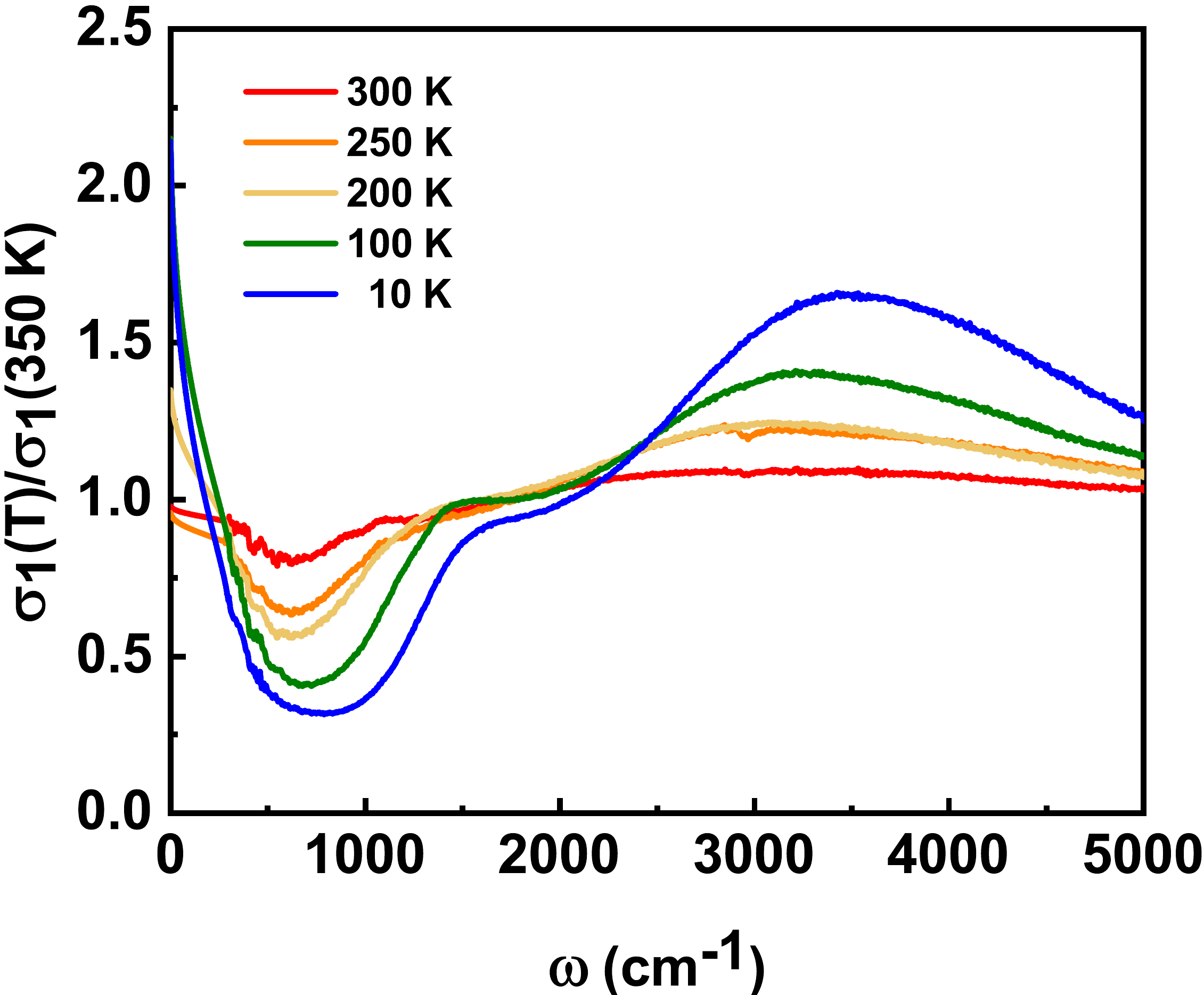}
    \includegraphics[width=6.8cm]{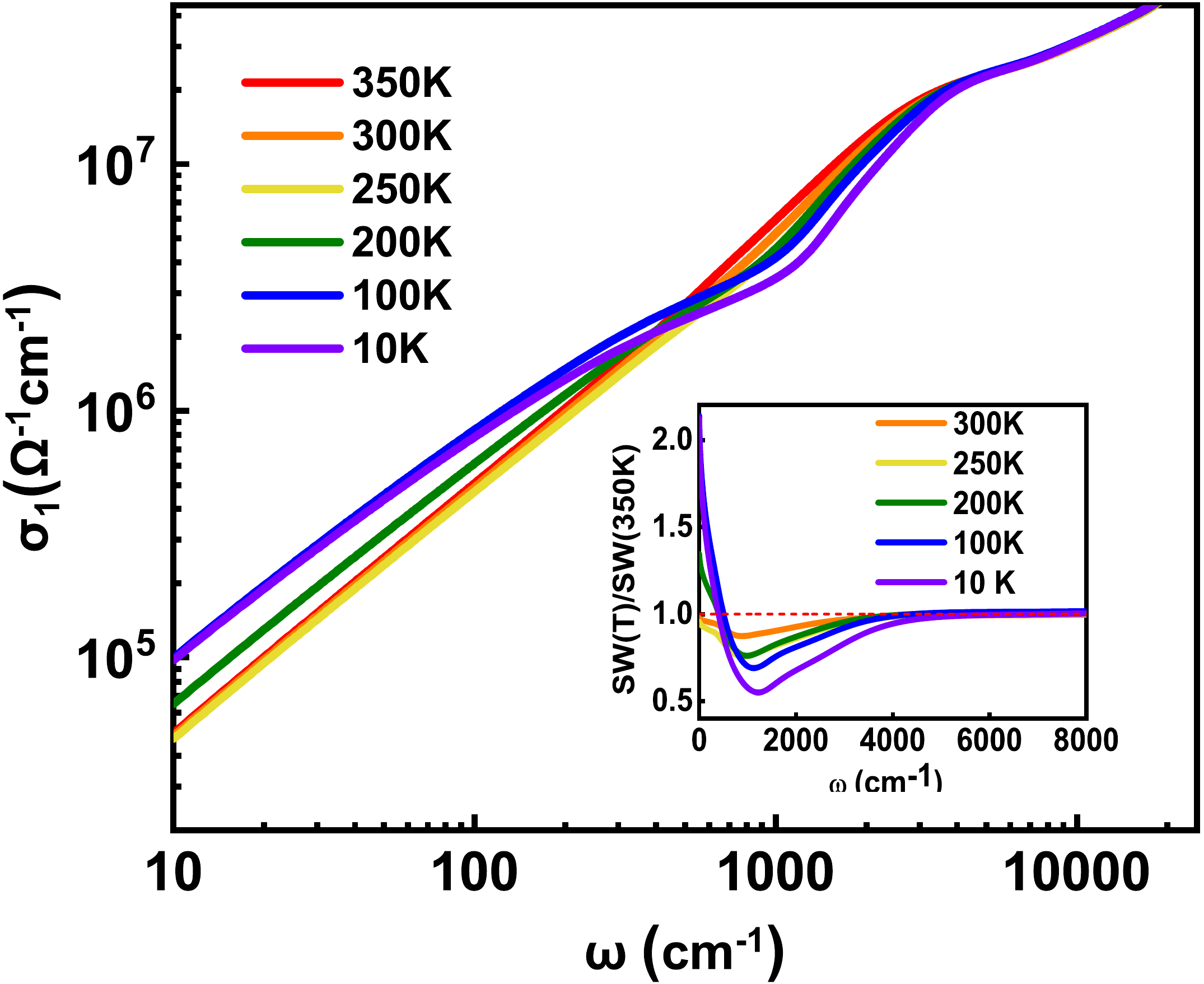}

  \caption{The frequency dependent (a) reflectivity R($\omega$) and (b) optical conductivity $\sigma_1(\omega)$  at several temperatures up to 10 000 \cm. The two insets show R($\omega$) and $\sigma_1(\omega)$ in an expanded range up to 50 000 \cm, respectively. (c) The renormalized optical conductivity as a function of frequency up to 5000 \cm. (d ) The spectral weight as a function of frequency. The inset shows renormalized SW at different temperatures.}\label{Fig:2}
\end{figure*}

The frequency dependent reflectivity $R(\omega)$ at several temperatures is shown in Fig.~\ref{Fig:2} (a). In the far-infrared (IR) region, metallic responses in the whole temperature range are clearly evidenced by the increasing of $R(\omega)$  with decreasing energy and its approaching to unity at zero frequency. Although HoSbTe was predicted to be a weak topological insulator, it could still be a metal since the Fermi level may not lie in the topological gaps and may pass through other trivial energy bands. A diamond-shaped Fermi surface (FS) around $\Gamma$ is indeed observed by ARPES experiments on HoSbTe\cite{Yue2020}. However this is somewhat inconsistent with the insulating behavior of $\rho(T)$ above 200 K, which will be discussed in the following.

In the mid-IR region, a platform-like feature can be clearly seen in $R(\omega)$ at 350 K, which is gradually suppressed at lower temperatures and ultimately replaced by two dip structures as indicated by the red arrows in Fig.~\ref{Fig:2}(a).
Such kind of low energy absorptions in $R(\omega)$ are usually linked to the loss of density of state at the Fermi level. The growing spectral weights of the two dips with temperature cooling further imply that they might be related to some kind of gap openings. For even higher frequencies, $R(\omega)$ decreases with increasing frequency up to 18 000 \cm and exhibits only mild temperature dependence. A few weak humps could be resolved in this frequency range.

 The evolution of the electronic structure could be more straightforwardly reflected by the optical conductivity $\sigma_1(\omega)$, which is directly related to the joint density of states for electron transitions from occupied to unoccupied states. As displayed in Fig.~\ref{Fig:2}(b), the direct current value of $\sigma_1(\omega\rightarrow0)$ first reduces slightly with decreasing temperature above 250K, then increases rapidly down to 10 K, roughly matching with the result of $\rho(T)$. The presence of the Drude components centered at zero frequency is another evidence of the metallic nature of HoSbTe.
In correspondence with the dip structures in $R(\omega)$, two extremely pronounced peaks could be identified in the optical conductivity $\sigma(\omega)$ at 10 K, where a crossover can be observed at around 2400 \cm between the two peaks. As temperature increases, the two individual peaks tend to get closer to each other and seem to turn into a very broad single peak above room temperature. Considering the two-dip structure in $R(\omega)$ can persist up to 300 K, we believe the broad peak in $\sigma_1(\omega)$ is actually composed by two independent peaks.

To better analyze the evolvement of the two peaks, we plot the renormalized optical conductivity spectra in Fig.\ref{Fig:2}(c). It is clearly shown that the energy scale of both peaks shift to higher frequency as temperature decreasing. Supposing they are originated from SOC gaps as revealed by the ARPES experiments or trivial interband transitions, the spectral weight of them should undergo a subtle decrease upon temperature cooling, due to the reduction of thermal excitations. While the lower energy peak is slightly suppressed as expected, the higher energy one gets much more pronounced when temperature decreases, suggesting a totally different origination.
We further present the spectral weight SW=$\int_{0}^{\omega}\sigma_1(\omega)d\omega$ as a function of frequency and temperature in Fig.\ref{Fig:2}(d).
At the low frequencies end, the SW increases as temperature cooling due to the narrowing of the Drude component. Then the higher temperature SW increases and surpasses the lower temperature value roughly above 700 \cm.
At even higher energies, SW at different temperatures almost merge together above 5000 \cm, complying with the conservation law. Additionally, the renormalized SW reaches a minimum at around 1200 \cm at 10 K, as can be seen in the inset of Fig.\ref{Fig:2}(d),
indicating the SW transfer from frequencies lower than 1200 \cm to higher energy region. Notably, these behaviors are similar with some CDW materials, such as CeTe$_3$\cite{Hu2011a}.

\begin{figure}
  \centering

   \includegraphics[width=2.5cm]{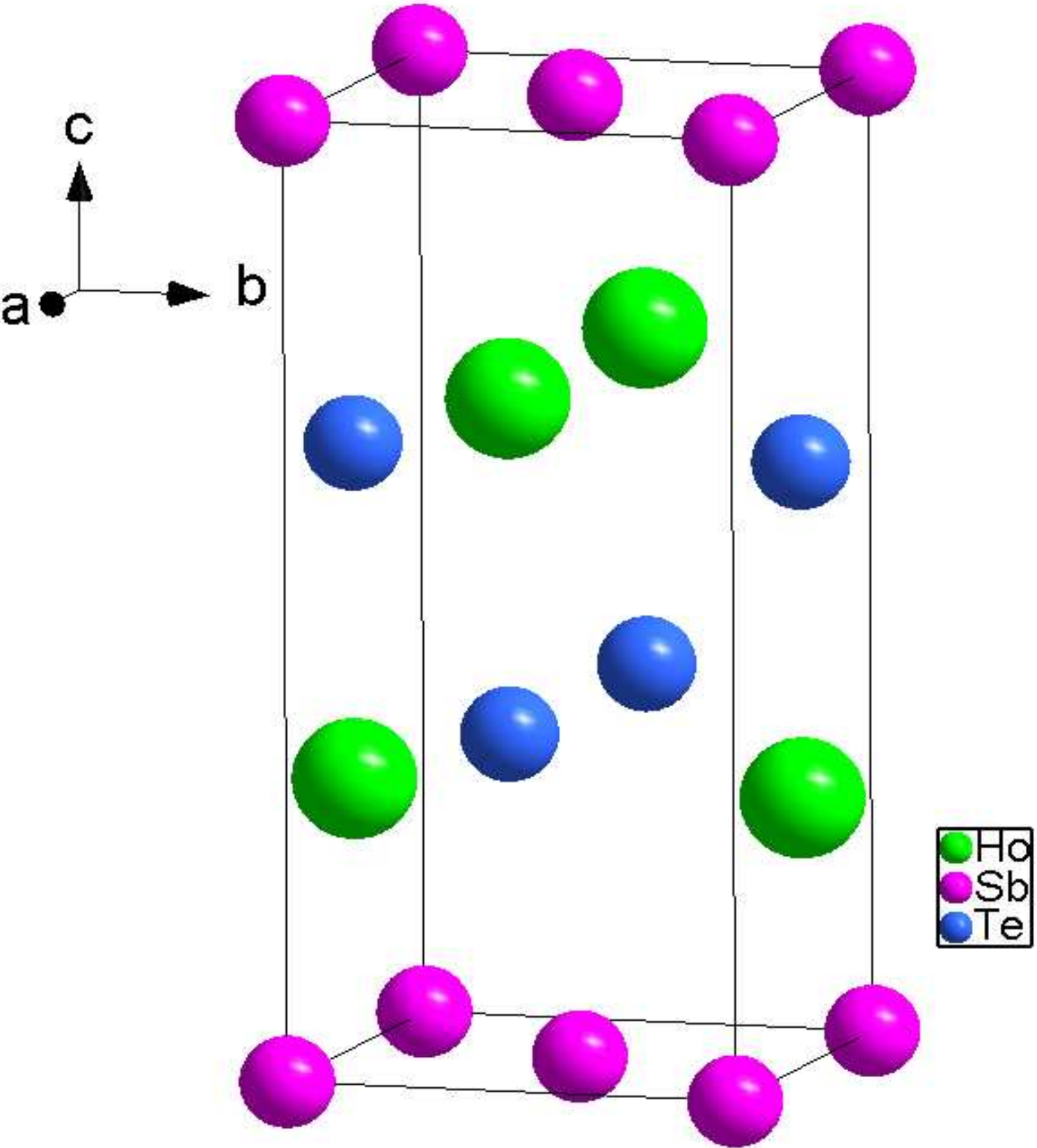}
      \includegraphics[width=5cm]{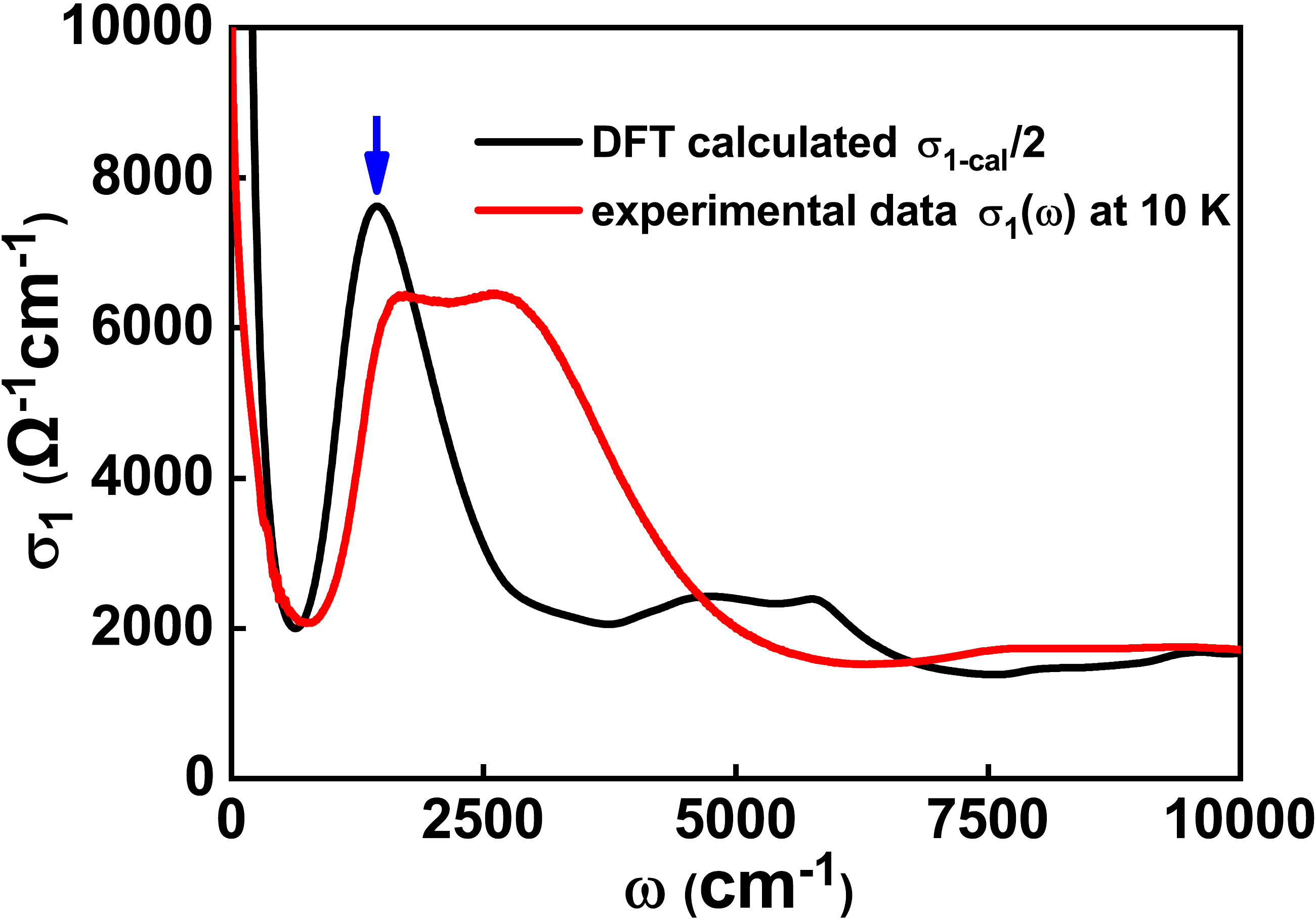}
   \includegraphics[width=2cm]{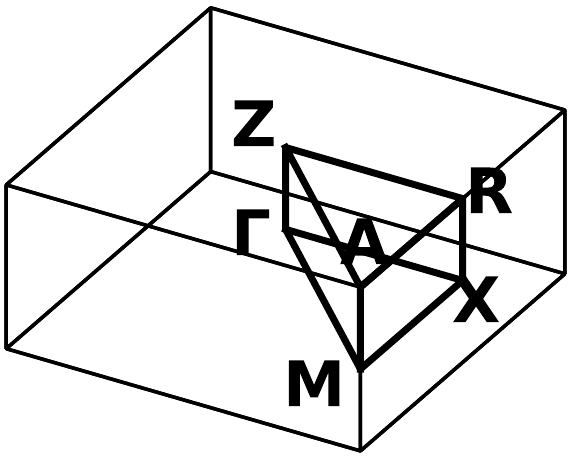}
  \includegraphics[width=5.5cm]{band.pdf}

  \caption{(a) Crystal structure of HoSbTe. A unit cell is indicated by black lines. (b) The calculated conductivity $\sigma_{1-cal}$ and experimental data at 10 K. (c) Brillouin zone of HoSbTe. (d) The calculated bulk band structure with orbital characters for HoSbTe. The light green arrow marks the optical interband transition that give rise to the mid-IR peak in the optical conductivity. Te 5$p$, Sb 5$p$ and Ho 4$f$ orbital characters are presented by red, blue, and green circle, respectively.}\label{Fig:band}
\end{figure}

 In order to unravel the individual mechanism of the two peaks, we performed first-principals calculations on the optical conductivity $\sigma_{1-cal}(\omega)$ and band structure of HoSbTe, as shown in Fig.\ref{Fig:band}(b) and (d) respectively. Surprisingly, there are some distinct differences between the calculated and experimental optical conductivity: the overall calculated spectra are much higher than the experimental data, partially due to the high-energy extrapolation of $R(\omega)$; $\sigma_{1-cal}(\omega)$  only shows a single peak in the mid-IR region at around 1450 \cm, which corresponds to interband transition across an SOC associated gap, as indicated by the light green arrow in Fig.\ref{Fig:band}(d). These disagreements clearly indicate a yet unknown reconstruction of the electronic structure.
Nevertheless, as the lowest energy interband transition,
the SOC gap in $\sigma_{1-cal}(\omega)$ locates very close to the lower energy peak in the experimental data, pointing to a similar origination.

\begin{figure}[htb]
  \includegraphics[width=7.5cm]{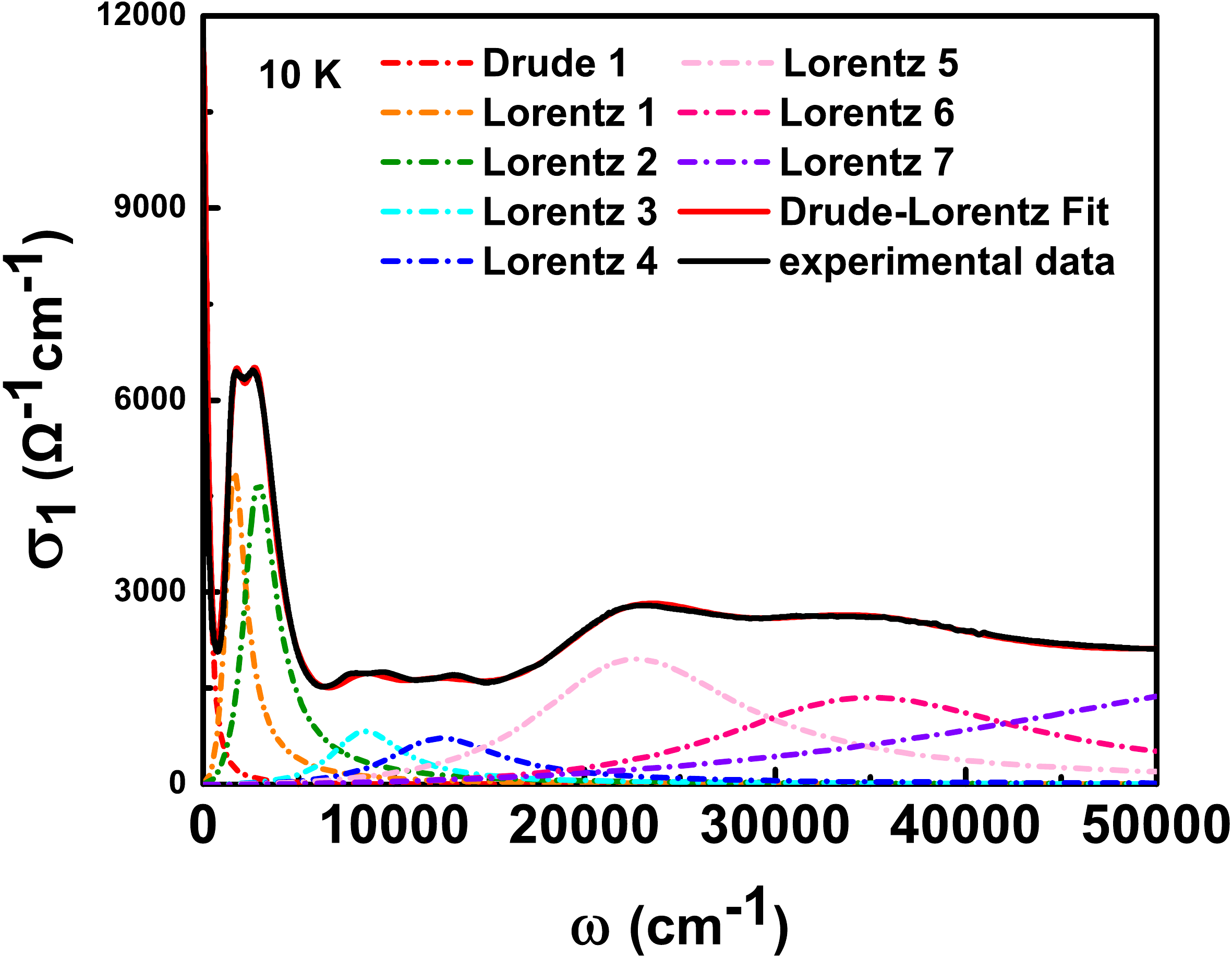}
  \caption{The Drude-Lorentz model fitting result of optical conductivity $\sigma_1(\omega)$ at 10 K, where one Drude term and seven Lorentz terms are used.}\label{Fig:DL}
\end{figure}

We further analyze the optical conductivity quantitatively by decomposing it into several different terms according to the Drude-Lorentz model:
 \begin{equation}\label{Eq:1}
  \sigma_1(\omega)=\frac{\omega_p^2}{4\pi}\frac{\gamma_D}{\omega^2+\gamma_D^2}+\sum_j\frac{S_{j}^2}{4\pi}\frac{\gamma_j\omega^2}{(\omega_j^2-\omega^2)^2+\omega^2\gamma_j^2},
\end{equation}
where $\omega_p$ and $\gamma_D$ are the plasma frequency and scattering rate of the itinerant carriers, while $\omega_j$, $\gamma_j$ and $S_j$ are the resonant frequency, damping and strength of the $j$th Lorentz oscillators, respectively. The first term at the right side of equation (\ref{Eq:1}) is the Drude component modeling intraband transitions of free electrons while the second Lorentz term describes interband transitions across energy gaps.
A presentative example of the fitting results of $\sigma_1(\omega)$ at 10 K are plotted in Fig.\ref{Fig:DL}, where one Durde and seven Lorentz terms are employed. Part of the fitting parameters are listed in Table \ref{TA:I}.

\begin{table}[h]
\scriptsize
\begin{tabular}{c|cccccccc}
  \hline
  \hline
  T & $\omega_P$ & $\gamma_D$ & $\omega_1$ & $\gamma_1$ & $S_1$ & $\omega_2$ & $\gamma_2$ & $S_2$ \\
  \hline
  10K & 11311(2)& 269 & 1647 & 1276(2) & 19455(16) & 2934 & 2173(1) & 24967 \\
  100K & 12417(2) & 305 & 1492 & 1224(1) & 19830(14) & 2700 & 2183(1) & 24163(14) \\
  200K & 12983(2) & 455 & 1405 & 1504(1) & 21539(15) & 2523 & 2177(1) & 22211(15) \\
  250K & 14698(7) & 752 & 1363 & 1465(2) & 18848(23) & 2432 & 2223 & 23560(16) \\
  300K & 19175 & 1246 & 1396 & 1730 & 18581 & 2335 & 2171 & 20089 \\
   & (46) & (6) & (1) & (6) & (83) & (1) & (1) & (39) \\
  350K & 18066 & 1077 & 1089 & 1938 & 18925 & 2032 & 2247 & 21267 \\
  & (1855) & (223) & (5) & (53) & (1900) & (2) & (5) & (135) \\
  \hline
  \hline
\end{tabular}
\caption{The fitting parameters at different temperatures. $\omega_P$ is the plasma frequency and $\gamma_D$=1/$\tau_D$ is the scattering rate of the Drude term. The center frequency $\omega_j$, width $\gamma_j$=1/$\tau_j$ and square root of the oscillator strength $S_j$ of the first two Lorentz components are listed. The error bars of all fitting parameters are included.}\label{TA:I}
\end{table}

\begin{figure}
\centering
 \subfigure{
 \begin{minipage}[b]{0.4\textwidth}
  \includegraphics[width=7.7cm]{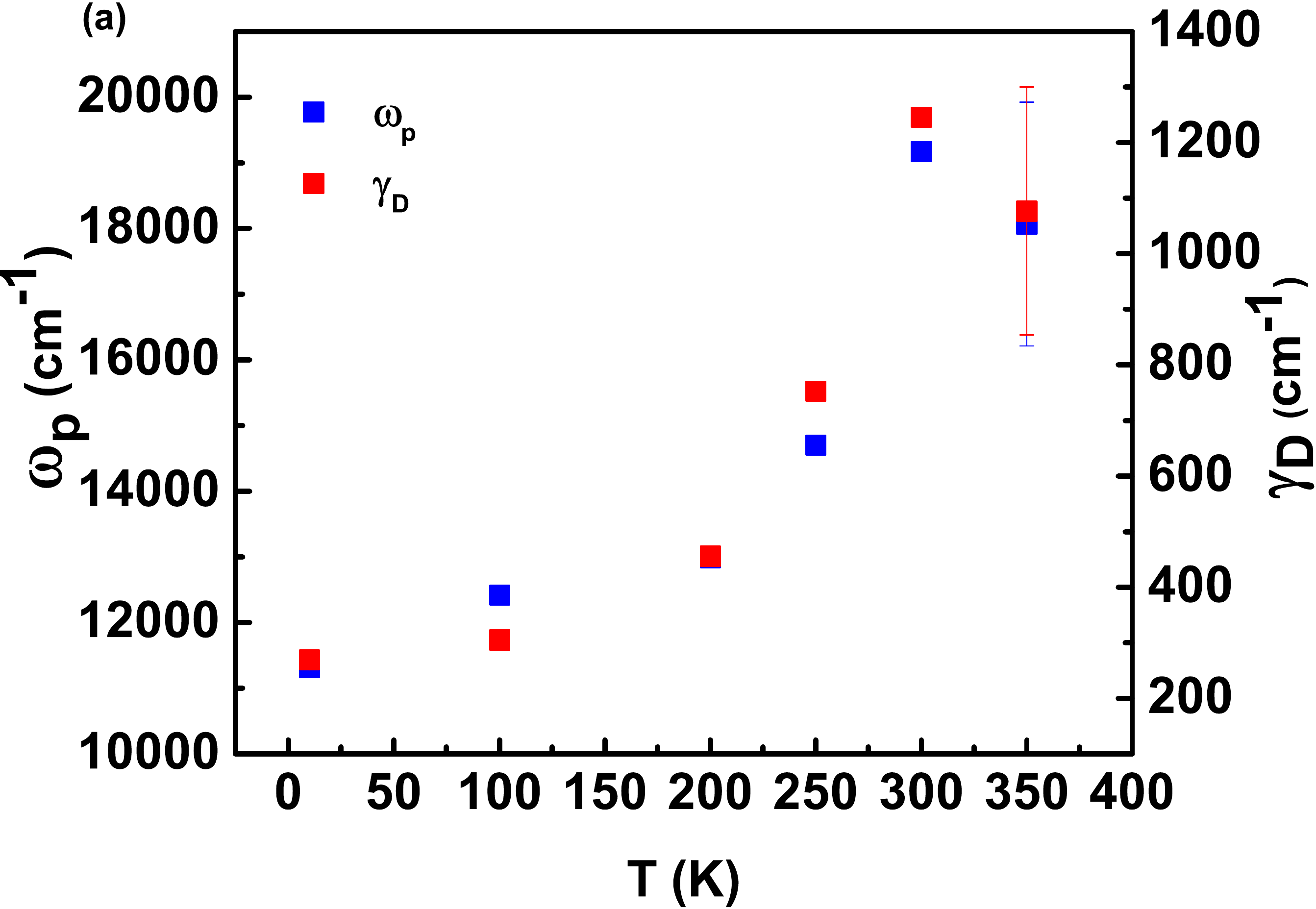}
  \includegraphics[width=6.85cm]{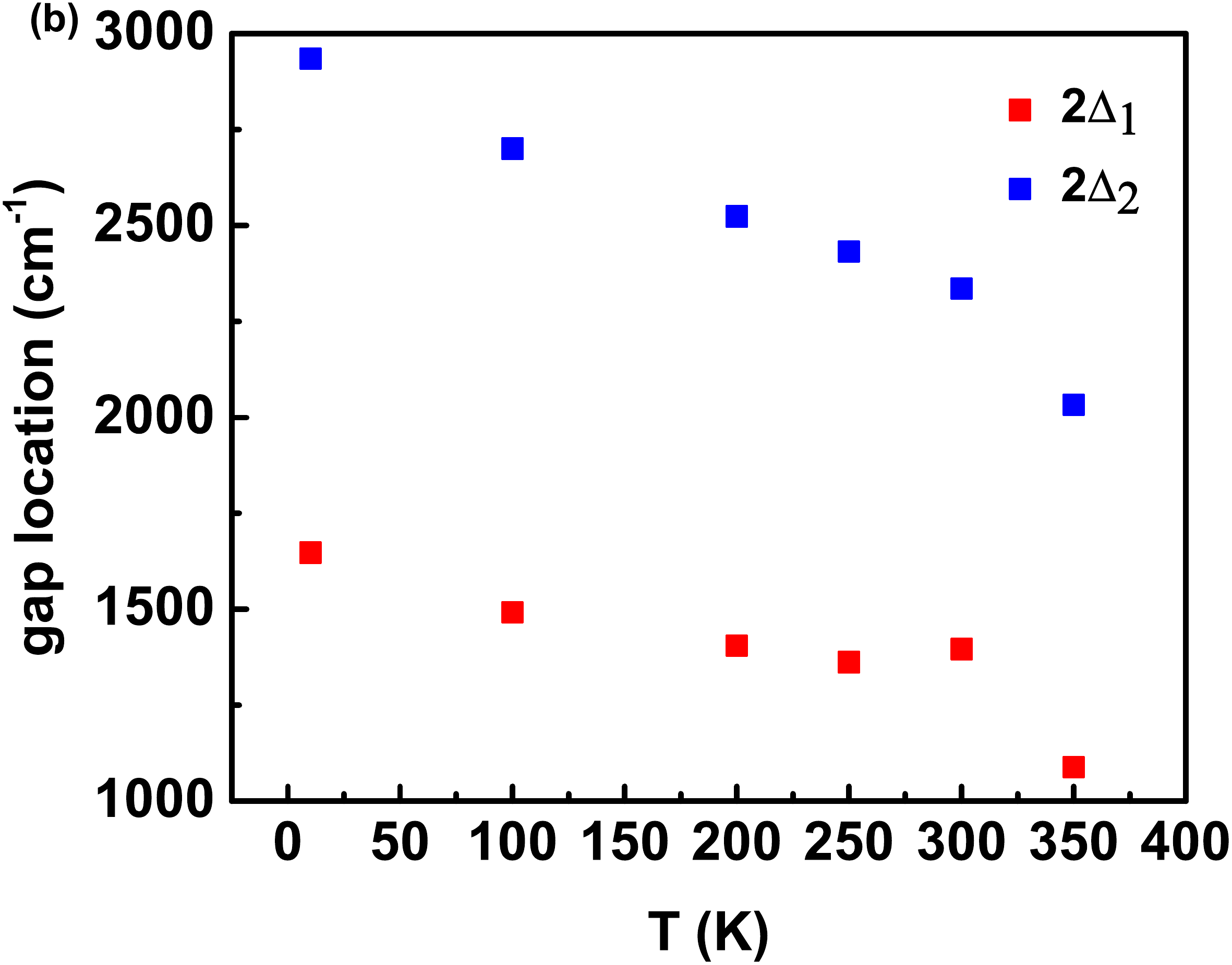}
 \end{minipage}
 }
  \caption{(a) The temperature dependent plasma frequency $\omega_p$ and scattering rate $\gamma_D$ . (b) The energy scales of the two gaps as a function of temperature.}\label{Fig:4}
\end{figure}

Among the fitting parameters, the plasma frequency $\omega_p$ is related to the itinerant carrier density $n$ and effective mass $m^*$ by $\omega_p^2=4\pi n e^2/m^*$.
The variation of $\omega_p$  with temperature is shown in Fig.\ref{Fig:4} (a), where $\omega_p$ decreases monotonically from 19 175 \cm at 300 K to 11 311 \cm at 10 K, indicating either a continuous losing of the free carrier density or a constant enhancement of the effective mass. It is worth to remark that the data of 350 K seems to exhibit a deviation from the above mentioned trend. This is because the Drude pattern of $\sigma_1(\omega)$ at 350 K is very broad and highly overlap with the first Lorentz term. Therefore, an extremely huge error bar is generated in the fitting procedure.
 The temperature dependent scattering rate $\gamma_D$, which is the width at half maximum of the Drude spectra, is
plotted in Fig.\ref{Fig:4} (a) as well. Notably, $\gamma_D$ evolves in a very similar fashion with $\omega_p$.
It is well known that the electrical conductivity of a material is proportional to its density of free carriers $n$ and relaxation time $\tau=1/\gamma_D$. Here for HoSbTe, $\tau$ obviously increases with lowering temperature, in favor of a  diminishing resistivity. As a consequence, how the free carrier density $n$ evolve with temperature is crucial to explain the hump feature in $\rho(T)$, which will be further elaborated later.

The two prominent peaks of $\sigma_1(\omega)$ in the mid-IR region could be well fitted by two Lorentz oscillators, corresponding to transitions across two different gaps. The obtained resonant frequencies, as displayed in Fig.\ref{Fig:4} (b), can be considered as the excitation gap energies, which are identified to be $2\Delta_1\simeq$ 204 meV and $2\Delta_2\simeq$ 364 meV at 10 K, respectively. In a previous report\cite{Yue2020}, ARPES measurements have revealed SOC gaps as large as hundreds of meV, which seems to be consistent with both $2\Delta_1$ and $2\Delta_2$.
As discussed above,
although this SOC scenario or trivial interband transitions can work for the lower energy gap, they can not explain the drastic decrease of $2\Delta_2$ to 252 meV at 350 K and the increase of its SW at lower temperatures. Taking the closeness between the calculated SOC gap and $2\Delta_1$ into consideration, we propose this lower energy gap to be an SOC gap, while the higher energy gap requires an alternative mechanism which will be discussed in the following.

In consideration of the exceptional large Sommerfeld coefficient $\gamma$\cite{Yang2020}, there is a possibility that the higher energy gap arises from the Kondo effect, in which case the interaction between located 4$f$ electrons and conduction bands would generate a mid-IR peak. The Drude spectral weight as well as the plasma frequency $\omega_p$ will be severely suppressed owing to the enhancement of effective mass $m^*$. This is in excellent agrement with our spectroscopic results. Even the drop of $\rho(T)$ below 200 K could be well justified because the initial localized $f$ electrons will be enclosed by the Fermi surface and contribute carrier density as a result of Kondo screening, which reduces the scattering rate of itinerant carriers at the same time. However, even if we adopt a constant carrier density, the enhancement of $m^*$ from 350 K to 10 K is estimated to be only about $m^*_{10K}/m^*_{350K}=\omega^2_{p(350K)}/\omega^2_{p(10K)}\simeq 2.55$, which disagrees violently with the huge Sommerfeld coefficient $\gamma$ = 382.2 mJ/mol$^{-1}$/K$^2$\cite{Yang2020}.
Besides, the sustaining of $2\Delta_2$ up to 350 K would indicate an even higher Kondo temperature $T_K$, below which the magnetic susceptibly is supposed to deviate from the Curie-Weiss law along with a downward bending resistivity. Such characters is obviously absent in our measurements. Hence we have ruled out the existence of Kondo effect in this system .

\begin{figure}
  \centering
  \includegraphics[width=7.5cm]{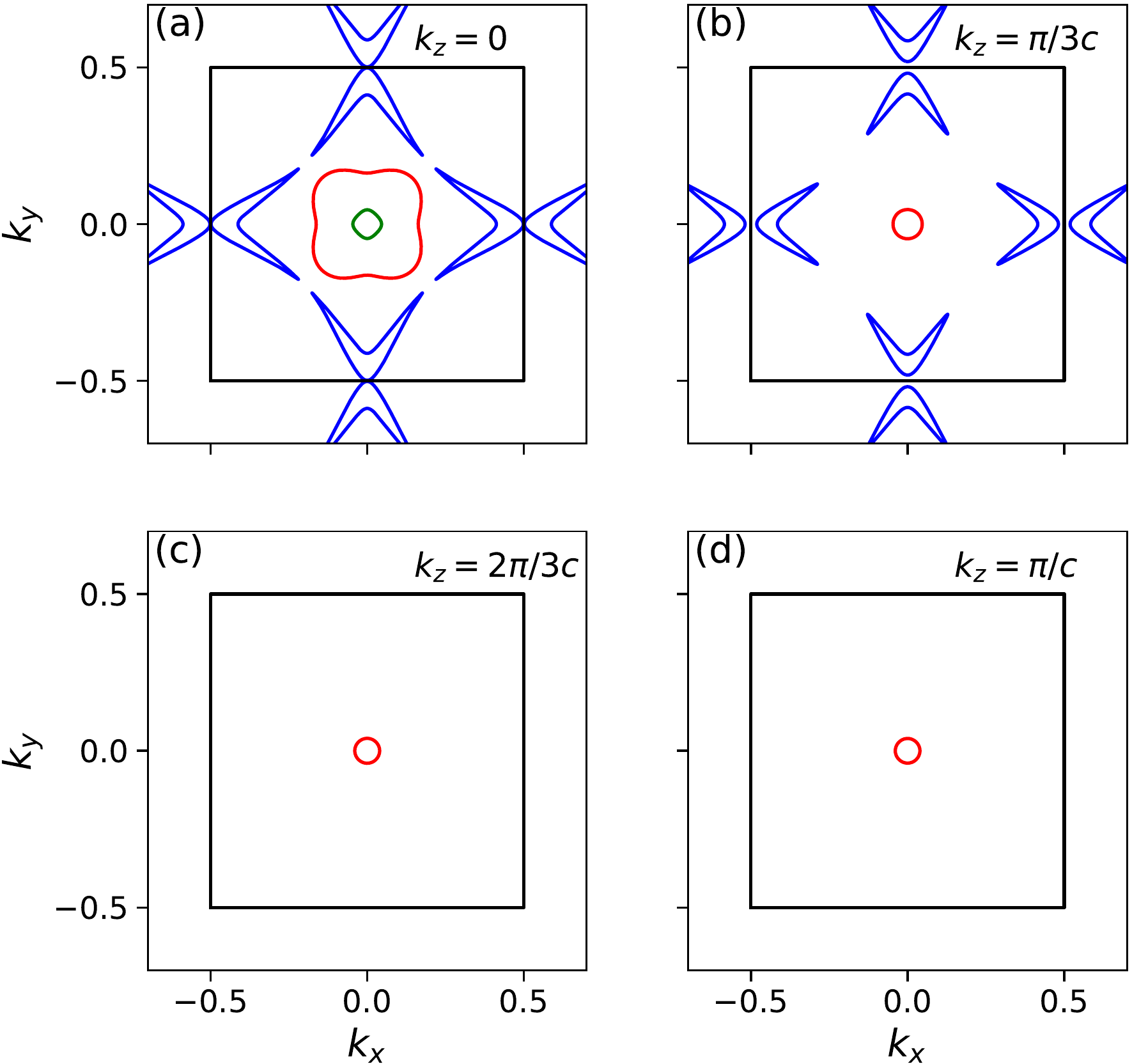}
  \caption{Fermi surface sections in (a) $kz$ = 0, (b) $kz$ = $\pi$/3c, (c) $kz$ = 2$\pi$/3c and (d) $kz$ = $\pi$/c plane of HoSbTe. Red, green and blue curves represent Fermi surfaces formed by different bands which are divided by the order of energy. The black lines mark the boundaries of Brillouin zone.}\label{Fig:FS}
\end{figure}

Another option for $2\Delta_2$ is a density-wave gap, since the transfer of spectral weight from the free electron Drude response to higher energy peaks agrees perfectly with the electrodynamic prediction of density-wave orderings with a case-I coherent factor.
Notably, a charge density wave gap about 0.3 eV is observed in the Dirac semimetal CeSbTe\cite{Li2021}, driven by electron-phonon coupling due to FS nesting. The FSs of CeSbTe are constituted by a small enclosed hole pocket at $\Gamma$ and two 2D rhombus-shaped sheets, the latter of which contain largely parallel segments that could be connected by a single wave
vector, in favor of FS nesting and thus the development of density wave orderings. For comparison, we plot the FS section of HoSbTe at several specified $kz$ planes, as shown in Fig.\ref{Fig:FS}. It also contains two rhombus-shape sheets at $kz$=0, but the two sheets are closer to each other and disconnected along the $\Gamma-M$ direction due to SOC gaps. When $kz$ increases, they gradually shrinks and eventually disappears as the SOC gaps shift closer to the Fermi level.
As a result, the two sheets are indistinguishable within the energy resolution limitation of ARPES measurements\cite{Yue2020}. In reality, however, it is most likely the nearly perfect nesting of these two sheets induces a density wave instability, and the corresponding single particle energy gap is captured by our infrared spectroscopy. As there are no additional  signatures in support of spin-density-wave instabilities, like AFM orderings above 10 K, we speculate that $2\Delta_2$ is actually a charge density wave gap. Since the two rhombus-shape are so close to each other, an extremely small nesting wave vector is expected.

Assuming that $m^*$ is temperature independent and equal to the electron rest mass $m_e$, $n$ is estimated to be about 3.62 $\times$ 10$^{21}$ $cm^{-3}$ at 350 K, which reduce to 1.42 $\times$ 10$^{21}$ $cm^{-3}$ at 10 K, meaning a 60.8$\%$ loss of the FS due to the gap enlargement. The residual FS at the lowest temperature manifests that the CDW gaps are partially opened.
As addressed above, the scattering rate $\gamma_D$ also decreases with temperature lowering. Consequently, the broad hump feature appeared in $\rho(T)$ is probably caused by the competition of the simultaneous reduction of $n$ and $\gamma_D$, which contribute oppositely to the resistivity. This phenomena is also observed in the three dimensional Dirac semimtal ZrTe$_5$\cite{Chen2015}.

Although we can principally ascertain the occurrence of a CDW ordering in HoSbTe, we have not been able to determine its phase transition temperature and ordering wave vector from our current results.
The large gap energies at 350 K suggest that the CDW transition takes place at a much higher temperature, which is inaccessible for our current measurement.
In the weak coupling mean field theory framework, the gap energy 2$\Delta$ = 3.52$k_B T_C$ for strict 1D systems, where $T_C$ is the phase transition temperature. By this relation the CDW $T_C$ of HoSbTe would be over 2000 K with gap value 2$\Delta_2=364$ meV. However $T_C$ will be severely suppressed when the effect of fluctuations is taken into account, especially in strongly anisotropic materials. Therefore the 2$\Delta$/$k_BT_C$ ratio is typically much larger than 3.52.
 For example, another Ho-based material HoTe$_3$ undergoes a CDW transition at 288 K with an energy gap of 380 meV\cite{HoTe3}, very close to the energy scale of $2\Delta_2$. We therefore conjecture that the transition temperature of HoSbTe is probably of the same order of magnitude with HoTe$_3$.


Finally we would like to compare HoSbTe with LaSbTe and CeSbTe. Although all of the three compounds bear strikingly similar FSs\cite{Yue2020,Wang2021,Li2021}, their physical properties are much different from each other. Our results on HoSbTe and previous reports on CeSbTe demonstrate that these two materials have CDW instabilities, while LaSbTe is reported to be a nodal line semimetal\cite{Wang2021} with no CDW observed by so far.
For CeSbTe the CDW gap is estimated to be about 0.3 eV with a transition temperature above 338 K and a weak Kondo effect is also identified\cite{Li2021}. We observe a CDW gap of about 364 meV in HoSbTe with no Kondo effect. This rich phase diagram supplies an ideal platform to investigate the influence of chemical pressure on the CDW ordering and its interplay with Kondo effect by substituting $Ln$SbTe with different lanthanide elements.


\section{conclusion}

In summary, we have investigated the charge carrier dynamics of a predicted topological insulator compound HoSbTe. The temperature dependent reflectivity and optical conductivity both exhibit metallic behaviors from 350 K to 10 K, with a decreasing carrier density and scattering rate, the competing contribution of which gives rise to the hump feature in resistivity. We identify an expected SOC gap of around 204 meV, and an additional temperature dependent gap whose value shifts from 364 meV at 10 K to 252 meV at 350 K. This enormous change along with its sizable increase of spectral weight upon cooling are clearly incompatible with SOC band gaps or trivial interband transitions. As Kondo hybridization is also excluded as the origin of this gap, due to a minor enhancement of the effective mass, we ultimately propose it to stem from a CDW ordering. This complex system together with its LnSbTe cousins provide a promising arena to explore the entanglement of CDW, Kondo effect and topological orders.

\begin{center}
\small{\textbf{ACKNOWLEDGMENTS}}
\end{center}

This work was supported by the National Science Foundation of China (No. 12074042),
the Young Scientists Fund of the National Natural Science Foundation of China (Grant No. 11704033), the National Key Research and Development Program of China (No. 2016YFA0302300),  and the Fundamental Research Funds for the Central Universities.

\bibliographystyle{apsrev4-1}
  \bibliography{HoSbTe}

\begin{thebibliography}{30}%
\makeatletter
\providecommand \@ifxundefined [1]{%
 \@ifx{#1\undefined}
}%
\providecommand \@ifnum [1]{%
 \ifnum #1\expandafter \@firstoftwo
 \else \expandafter \@secondoftwo
 \fi
}%
\providecommand \@ifx [1]{%
 \ifx #1\expandafter \@firstoftwo
 \else \expandafter \@secondoftwo
 \fi
}%
\providecommand \natexlab [1]{#1}%
\providecommand \enquote  [1]{``#1''}%
\providecommand \bibnamefont  [1]{#1}%
\providecommand \bibfnamefont [1]{#1}%
\providecommand \citenamefont [1]{#1}%
\providecommand \href@noop [0]{\@secondoftwo}%
\providecommand \href [0]{\begingroup \@sanitize@url \@href}%
\providecommand \@href[1]{\@@startlink{#1}\@@href}%
\providecommand \@@href[1]{\endgroup#1\@@endlink}%
\providecommand \@sanitize@url [0]{\catcode `\\12\catcode `\$12\catcode
  `\&12\catcode `\#12\catcode `\^12\catcode `\_12\catcode `\%12\relax}%
\providecommand \@@startlink[1]{}%
\providecommand \@@endlink[0]{}%
\providecommand \url  [0]{\begingroup\@sanitize@url \@url }%
\providecommand \@url [1]{\endgroup\@href {#1}{\urlprefix }}%
\providecommand \urlprefix  [0]{URL }%
\providecommand \Eprint [0]{\href }%
\providecommand \doibase [0]{http://dx.doi.org/}%
\providecommand \selectlanguage [0]{\@gobble}%
\providecommand \bibinfo  [0]{\@secondoftwo}%
\providecommand \bibfield  [0]{\@secondoftwo}%
\providecommand \translation [1]{[#1]}%
\providecommand \BibitemOpen [0]{}%
\providecommand \bibitemStop [0]{}%
\providecommand \bibitemNoStop [0]{.\EOS\space}%
\providecommand \EOS [0]{\spacefactor3000\relax}%
\providecommand \BibitemShut  [1]{\csname bibitem#1\endcsname}%
\let\auto@bib@innerbib\@empty
\bibitem [{\citenamefont {Qi}\ and\ \citenamefont
  {Zhang}(2011)}]{RevModPhys.83.1057}%
  \BibitemOpen
  \bibfield  {author} {\bibinfo {author} {\bibfnamefont {X.-L.}\ \bibnamefont
  {Qi}}\ and\ \bibinfo {author} {\bibfnamefont {S.-C.}\ \bibnamefont {Zhang}},\
  }\href {\doibase 10.1103/RevModPhys.83.1057} {\bibfield  {journal} {\bibinfo
  {journal} {Rev. Mod. Phys.}\ }\textbf {\bibinfo {volume} {83}},\ \bibinfo
  {pages} {1057} (\bibinfo {year} {2011})}\BibitemShut {NoStop}%
\bibitem [{\citenamefont {Hasan}\ and\ \citenamefont
  {Kane}(2010)}]{RevModPhys.82.3045}%
  \BibitemOpen
  \bibfield  {author} {\bibinfo {author} {\bibfnamefont {M.~Z.}\ \bibnamefont
  {Hasan}}\ and\ \bibinfo {author} {\bibfnamefont {C.~L.}\ \bibnamefont
  {Kane}},\ }\href {\doibase 10.1103/RevModPhys.82.3045} {\bibfield  {journal}
  {\bibinfo  {journal} {Rev. Mod. Phys.}\ }\textbf {\bibinfo {volume} {82}},\
  \bibinfo {pages} {3045} (\bibinfo {year} {2010})}\BibitemShut {NoStop}%
\bibitem [{\citenamefont {Nayak}\ \emph {et~al.}(2008)\citenamefont {Nayak},
  \citenamefont {Simon}, \citenamefont {Stern}, \citenamefont {Freedman},\ and\
  \citenamefont {Das~Sarma}}]{RevModPhys.80.1083}%
  \BibitemOpen
  \bibfield  {author} {\bibinfo {author} {\bibfnamefont {C.}~\bibnamefont
  {Nayak}}, \bibinfo {author} {\bibfnamefont {S.~H.}\ \bibnamefont {Simon}},
  \bibinfo {author} {\bibfnamefont {A.}~\bibnamefont {Stern}}, \bibinfo
  {author} {\bibfnamefont {M.}~\bibnamefont {Freedman}}, \ and\ \bibinfo
  {author} {\bibfnamefont {S.}~\bibnamefont {Das~Sarma}},\ }\href {\doibase
  10.1103/RevModPhys.80.1083} {\bibfield  {journal} {\bibinfo  {journal} {Rev.
  Mod. Phys.}\ }\textbf {\bibinfo {volume} {80}},\ \bibinfo {pages} {1083}
  (\bibinfo {year} {2008})}\BibitemShut {NoStop}%
\bibitem [{\citenamefont {Pesin}\ and\ \citenamefont
  {Balents}(2010)}]{Pesin2010}%
  \BibitemOpen
  \bibfield  {author} {\bibinfo {author} {\bibfnamefont {D.}~\bibnamefont
  {Pesin}}\ and\ \bibinfo {author} {\bibfnamefont {L.}~\bibnamefont
  {Balents}},\ }\href {\doibase 10.1038/nphys1606} {\bibfield  {journal}
  {\bibinfo  {journal} {Nature Physics}\ }\textbf {\bibinfo {volume} {6}},\
  \bibinfo {pages} {376} (\bibinfo {year} {2010})}\BibitemShut {NoStop}%
\bibitem [{\citenamefont {Deng}\ \emph {et~al.}(2017)\citenamefont {Deng},
  \citenamefont {Luo}, \citenamefont {Wang}, \citenamefont {Sheng},\ and\
  \citenamefont {Xing}}]{Deng2017}%
  \BibitemOpen
  \bibfield  {author} {\bibinfo {author} {\bibfnamefont {M.-X.}\ \bibnamefont
  {Deng}}, \bibinfo {author} {\bibfnamefont {W.}~\bibnamefont {Luo}}, \bibinfo
  {author} {\bibfnamefont {R.-Q.}\ \bibnamefont {Wang}}, \bibinfo {author}
  {\bibfnamefont {L.}~\bibnamefont {Sheng}}, \ and\ \bibinfo {author}
  {\bibfnamefont {D.~Y.}\ \bibnamefont {Xing}},\ }\href {\doibase
  10.1103/PhysRevB.96.155141} {\bibfield  {journal} {\bibinfo  {journal} {Phys.
  Rev. B}\ }\textbf {\bibinfo {volume} {96}},\ \bibinfo {pages} {155141}
  (\bibinfo {year} {2017})}\BibitemShut {NoStop}%
\bibitem [{\citenamefont {Schoop}\ \emph {et~al.}(2018)\citenamefont {Schoop},
  \citenamefont {Topp}, \citenamefont {Lippmann}, \citenamefont {Orlandi},
  \citenamefont {M{\"{u}}chler}, \citenamefont {Vergniory}, \citenamefont
  {Sun}, \citenamefont {Rost}, \citenamefont {Duppel}, \citenamefont
  {Krivenkov}, \citenamefont {Sheoran}, \citenamefont {Manuel}, \citenamefont
  {Varykhalov}, \citenamefont {Yan}, \citenamefont {Kremer}, \citenamefont
  {Ast},\ and\ \citenamefont {Lotsch}}]{Schoop2018}%
  \BibitemOpen
  \bibfield  {author} {\bibinfo {author} {\bibfnamefont {L.~M.}\ \bibnamefont
  {Schoop}}, \bibinfo {author} {\bibfnamefont {A.}~\bibnamefont {Topp}},
  \bibinfo {author} {\bibfnamefont {J.}~\bibnamefont {Lippmann}}, \bibinfo
  {author} {\bibfnamefont {F.}~\bibnamefont {Orlandi}}, \bibinfo {author}
  {\bibfnamefont {L.}~\bibnamefont {M{\"{u}}chler}}, \bibinfo {author}
  {\bibfnamefont {M.~G.}\ \bibnamefont {Vergniory}}, \bibinfo {author}
  {\bibfnamefont {Y.}~\bibnamefont {Sun}}, \bibinfo {author} {\bibfnamefont
  {A.~W.}\ \bibnamefont {Rost}}, \bibinfo {author} {\bibfnamefont
  {V.}~\bibnamefont {Duppel}}, \bibinfo {author} {\bibfnamefont
  {M.}~\bibnamefont {Krivenkov}}, \bibinfo {author} {\bibfnamefont
  {S.}~\bibnamefont {Sheoran}}, \bibinfo {author} {\bibfnamefont
  {P.}~\bibnamefont {Manuel}}, \bibinfo {author} {\bibfnamefont
  {A.}~\bibnamefont {Varykhalov}}, \bibinfo {author} {\bibfnamefont
  {B.}~\bibnamefont {Yan}}, \bibinfo {author} {\bibfnamefont {R.~K.}\
  \bibnamefont {Kremer}}, \bibinfo {author} {\bibfnamefont {C.~R.}\
  \bibnamefont {Ast}}, \ and\ \bibinfo {author} {\bibfnamefont {B.~V.}\
  \bibnamefont {Lotsch}},\ }\href@noop {} {\bibfield  {journal} {\bibinfo
  {journal} {Science Advances}\ }\textbf {\bibinfo {volume} {4}},\ \bibinfo
  {pages} {eaar2317} (\bibinfo {year} {2018})}\BibitemShut {NoStop}%
\bibitem [{\citenamefont {Peters}\ \emph {et~al.}(2018)\citenamefont {Peters},
  \citenamefont {Yoshida},\ and\ \citenamefont {Kawakami}}]{Peters2018}%
  \BibitemOpen
  \bibfield  {author} {\bibinfo {author} {\bibfnamefont {R.}~\bibnamefont
  {Peters}}, \bibinfo {author} {\bibfnamefont {T.}~\bibnamefont {Yoshida}}, \
  and\ \bibinfo {author} {\bibfnamefont {N.}~\bibnamefont {Kawakami}},\ }\href
  {\doibase 10.1103/PhysRevB.98.075104} {\bibfield  {journal} {\bibinfo
  {journal} {Phys. Rev. B}\ }\textbf {\bibinfo {volume} {98}},\ \bibinfo
  {pages} {075104} (\bibinfo {year} {2018})}\BibitemShut {NoStop}%
\bibitem [{\citenamefont {Fu}\ and\ \citenamefont {Kane}(2008)}]{majorana}%
  \BibitemOpen
  \bibfield  {author} {\bibinfo {author} {\bibfnamefont {L.}~\bibnamefont
  {Fu}}\ and\ \bibinfo {author} {\bibfnamefont {C.~L.}\ \bibnamefont {Kane}},\
  }\href {\doibase 10.1103/PhysRevLett.100.096407} {\bibfield  {journal}
  {\bibinfo  {journal} {Phys. Rev. Lett.}\ }\textbf {\bibinfo {volume} {100}},\
  \bibinfo {pages} {096407} (\bibinfo {year} {2008})}\BibitemShut {NoStop}%
\bibitem [{\citenamefont {Wang}\ and\ \citenamefont {Zhang}(2013)}]{Wang2013}%
  \BibitemOpen
  \bibfield  {author} {\bibinfo {author} {\bibfnamefont {Z.}~\bibnamefont
  {Wang}}\ and\ \bibinfo {author} {\bibfnamefont {S.-c.}\ \bibnamefont
  {Zhang}},\ }\href {\doibase 10.1103/PhysRevB.87.161107} {\bibfield  {journal}
  {\bibinfo  {journal} {Physical Review B}\ }\textbf {\bibinfo {volume} {87}},\
  \bibinfo {pages} {161107(R)} (\bibinfo {year} {2013})}\BibitemShut {NoStop}%
\bibitem [{\citenamefont {Xu}\ \emph {et~al.}(2015)\citenamefont {Xu},
  \citenamefont {Song}, \citenamefont {Nie}, \citenamefont {Weng},
  \citenamefont {Fang},\ and\ \citenamefont {Dai}}]{Xu2015}%
  \BibitemOpen
  \bibfield  {author} {\bibinfo {author} {\bibfnamefont {Q.}~\bibnamefont
  {Xu}}, \bibinfo {author} {\bibfnamefont {Z.}~\bibnamefont {Song}}, \bibinfo
  {author} {\bibfnamefont {S.}~\bibnamefont {Nie}}, \bibinfo {author}
  {\bibfnamefont {H.}~\bibnamefont {Weng}}, \bibinfo {author} {\bibfnamefont
  {Z.}~\bibnamefont {Fang}}, \ and\ \bibinfo {author} {\bibfnamefont
  {X.}~\bibnamefont {Dai}},\ }\href {\doibase 10.1103/PhysRevB.92.205310}
  {\bibfield  {journal} {\bibinfo  {journal} {Physical Review B}\ }\textbf
  {\bibinfo {volume} {92}},\ \bibinfo {pages} {205310} (\bibinfo {year}
  {2015})}\BibitemShut {NoStop}%
\bibitem [{\citenamefont {Guo}\ \emph {et~al.}(2019)\citenamefont {Guo},
  \citenamefont {Chen}, \citenamefont {Chen}, \citenamefont {Chen},
  \citenamefont {Zhang}, \citenamefont {Gao}, \citenamefont {Yang},
  \citenamefont {Li}, \citenamefont {Zhao}, \citenamefont {Dong},\ and\
  \citenamefont {Zheng}}]{DNL-ZrGeSe}%
  \BibitemOpen
  \bibfield  {author} {\bibinfo {author} {\bibfnamefont {L.}~\bibnamefont
  {Guo}}, \bibinfo {author} {\bibfnamefont {T.-W.}\ \bibnamefont {Chen}},
  \bibinfo {author} {\bibfnamefont {C.}~\bibnamefont {Chen}}, \bibinfo {author}
  {\bibfnamefont {L.}~\bibnamefont {Chen}}, \bibinfo {author} {\bibfnamefont
  {Y.}~\bibnamefont {Zhang}}, \bibinfo {author} {\bibfnamefont {G.-Y.}\
  \bibnamefont {Gao}}, \bibinfo {author} {\bibfnamefont {J.}~\bibnamefont
  {Yang}}, \bibinfo {author} {\bibfnamefont {X.-G.}\ \bibnamefont {Li}},
  \bibinfo {author} {\bibfnamefont {W.-Y.}\ \bibnamefont {Zhao}}, \bibinfo
  {author} {\bibfnamefont {S.}~\bibnamefont {Dong}}, \ and\ \bibinfo {author}
  {\bibfnamefont {R.-K.}\ \bibnamefont {Zheng}},\ }\href {\doibase
  10.1021/acsaelm.9b00061} {\bibfield  {journal} {\bibinfo  {journal} {ACS
  Applied Electronic Materials}\ }\textbf {\bibinfo {volume} {1}},\ \bibinfo
  {pages} {869} (\bibinfo {year} {2019})}\BibitemShut {NoStop}%
\bibitem [{\citenamefont {Liu}\ \emph {et~al.}(0)\citenamefont {Liu},
  \citenamefont {Yue}, \citenamefont {Erohin}, \citenamefont {Zhu},
  \citenamefont {Joshy}, \citenamefont {Liu}, \citenamefont {Sanchez},
  \citenamefont {Graf}, \citenamefont {Sorokin}, \citenamefont {Mao},
  \citenamefont {Hu},\ and\ \citenamefont {Wei}}]{DNL-ZrSiSe}%
  \BibitemOpen
  \bibfield  {author} {\bibinfo {author} {\bibfnamefont {X.}~\bibnamefont
  {Liu}}, \bibinfo {author} {\bibfnamefont {C.}~\bibnamefont {Yue}}, \bibinfo
  {author} {\bibfnamefont {S.~V.}\ \bibnamefont {Erohin}}, \bibinfo {author}
  {\bibfnamefont {Y.}~\bibnamefont {Zhu}}, \bibinfo {author} {\bibfnamefont
  {A.}~\bibnamefont {Joshy}}, \bibinfo {author} {\bibfnamefont
  {J.}~\bibnamefont {Liu}}, \bibinfo {author} {\bibfnamefont {A.~M.}\
  \bibnamefont {Sanchez}}, \bibinfo {author} {\bibfnamefont {D.}~\bibnamefont
  {Graf}}, \bibinfo {author} {\bibfnamefont {P.~B.}\ \bibnamefont {Sorokin}},
  \bibinfo {author} {\bibfnamefont {Z.}~\bibnamefont {Mao}}, \bibinfo {author}
  {\bibfnamefont {J.}~\bibnamefont {Hu}}, \ and\ \bibinfo {author}
  {\bibfnamefont {J.}~\bibnamefont {Wei}},\ }\href {\doibase
  10.1021/acs.nanolett.0c04946} {\bibfield  {journal} {\bibinfo  {journal}
  {Nano Letters}\ }\textbf {\bibinfo {volume} {0}},\ \bibinfo {pages} {null}
  (\bibinfo {year} {0})}\BibitemShut {NoStop}%
\bibitem [{\citenamefont {Wang}\ \emph {et~al.}(2021)\citenamefont {Wang},
  \citenamefont {Qian}, \citenamefont {Yang}, \citenamefont {Chen},
  \citenamefont {Li}, \citenamefont {Tan}, \citenamefont {Cai}, \citenamefont
  {Zhao}, \citenamefont {Gao}, \citenamefont {Feng}, \citenamefont {Kumar},
  \citenamefont {Schwier}, \citenamefont {Zhao}, \citenamefont {Weng},
  \citenamefont {Shi}, \citenamefont {Wang}, \citenamefont {Song},
  \citenamefont {Huang}, \citenamefont {Shimada}, \citenamefont {Xu},
  \citenamefont {Zhou},\ and\ \citenamefont {Liu}}]{Wang2021}%
  \BibitemOpen
  \bibfield  {author} {\bibinfo {author} {\bibfnamefont {Y.}~\bibnamefont
  {Wang}}, \bibinfo {author} {\bibfnamefont {Y.}~\bibnamefont {Qian}}, \bibinfo
  {author} {\bibfnamefont {M.}~\bibnamefont {Yang}}, \bibinfo {author}
  {\bibfnamefont {H.}~\bibnamefont {Chen}}, \bibinfo {author} {\bibfnamefont
  {C.}~\bibnamefont {Li}}, \bibinfo {author} {\bibfnamefont {Z.}~\bibnamefont
  {Tan}}, \bibinfo {author} {\bibfnamefont {Y.}~\bibnamefont {Cai}}, \bibinfo
  {author} {\bibfnamefont {W.}~\bibnamefont {Zhao}}, \bibinfo {author}
  {\bibfnamefont {S.}~\bibnamefont {Gao}}, \bibinfo {author} {\bibfnamefont
  {Y.}~\bibnamefont {Feng}}, \bibinfo {author} {\bibfnamefont {S.}~\bibnamefont
  {Kumar}}, \bibinfo {author} {\bibfnamefont {E.~F.}\ \bibnamefont {Schwier}},
  \bibinfo {author} {\bibfnamefont {L.}~\bibnamefont {Zhao}}, \bibinfo {author}
  {\bibfnamefont {H.}~\bibnamefont {Weng}}, \bibinfo {author} {\bibfnamefont
  {Y.}~\bibnamefont {Shi}}, \bibinfo {author} {\bibfnamefont {G.}~\bibnamefont
  {Wang}}, \bibinfo {author} {\bibfnamefont {Y.}~\bibnamefont {Song}}, \bibinfo
  {author} {\bibfnamefont {Y.}~\bibnamefont {Huang}}, \bibinfo {author}
  {\bibfnamefont {K.}~\bibnamefont {Shimada}}, \bibinfo {author} {\bibfnamefont
  {Z.}~\bibnamefont {Xu}}, \bibinfo {author} {\bibfnamefont {X.~J.}\
  \bibnamefont {Zhou}}, \ and\ \bibinfo {author} {\bibfnamefont
  {G.}~\bibnamefont {Liu}},\ }\href {\doibase 10.1103/PhysRevB.103.125131}
  {\bibfield  {journal} {\bibinfo  {journal} {Physical Review B}\ }\textbf
  {\bibinfo {volume} {103}},\ \bibinfo {pages} {125131} (\bibinfo {year}
  {2021})}\BibitemShut {NoStop}%
\bibitem [{\citenamefont {Yen}\ \emph {et~al.}(2019)\citenamefont {Yen},
  \citenamefont {Chiu}, \citenamefont {Lin}, \citenamefont {Sankar},
  \citenamefont {Chou}, \citenamefont {Chuang},\ and\ \citenamefont
  {Guo}}]{DNL-ZrGeTe}%
  \BibitemOpen
  \bibfield  {author} {\bibinfo {author} {\bibfnamefont {Y.}~\bibnamefont
  {Yen}}, \bibinfo {author} {\bibfnamefont {C.-L.}\ \bibnamefont {Chiu}},
  \bibinfo {author} {\bibfnamefont {P.-H.}\ \bibnamefont {Lin}}, \bibinfo
  {author} {\bibfnamefont {R.}~\bibnamefont {Sankar}}, \bibinfo {author}
  {\bibfnamefont {F.}~\bibnamefont {Chou}}, \bibinfo {author} {\bibfnamefont
  {T.-M.}\ \bibnamefont {Chuang}}, \ and\ \bibinfo {author} {\bibfnamefont
  {G.-Y.}\ \bibnamefont {Guo}},\ }\href@noop {} {\enquote {\bibinfo {title}
  {Dirac nodal line and rashba splitting surface states in nonsymmorphic
  zrgete},}\ } (\bibinfo {year} {2019}),\ \Eprint
  {http://arxiv.org/abs/1912.07002} {arXiv:1912.07002} \BibitemShut {NoStop}%
\bibitem [{\citenamefont {Nakamura}\ \emph {et~al.}(2019)\citenamefont
  {Nakamura}, \citenamefont {Souma}, \citenamefont {Wang}, \citenamefont
  {Yamauchi}, \citenamefont {Takane}, \citenamefont {Oinuma}, \citenamefont
  {Nakayama}, \citenamefont {Horiba}, \citenamefont {Kumigashira},
  \citenamefont {Oguchi}, \citenamefont {Takahashi}, \citenamefont {Ando},\
  and\ \citenamefont {Sato}}]{DNL-ZrGeXc}%
  \BibitemOpen
  \bibfield  {author} {\bibinfo {author} {\bibfnamefont {T.}~\bibnamefont
  {Nakamura}}, \bibinfo {author} {\bibfnamefont {S.}~\bibnamefont {Souma}},
  \bibinfo {author} {\bibfnamefont {Z.}~\bibnamefont {Wang}}, \bibinfo {author}
  {\bibfnamefont {K.}~\bibnamefont {Yamauchi}}, \bibinfo {author}
  {\bibfnamefont {D.}~\bibnamefont {Takane}}, \bibinfo {author} {\bibfnamefont
  {H.}~\bibnamefont {Oinuma}}, \bibinfo {author} {\bibfnamefont
  {K.}~\bibnamefont {Nakayama}}, \bibinfo {author} {\bibfnamefont
  {K.}~\bibnamefont {Horiba}}, \bibinfo {author} {\bibfnamefont
  {H.}~\bibnamefont {Kumigashira}}, \bibinfo {author} {\bibfnamefont
  {T.}~\bibnamefont {Oguchi}}, \bibinfo {author} {\bibfnamefont
  {T.}~\bibnamefont {Takahashi}}, \bibinfo {author} {\bibfnamefont
  {Y.}~\bibnamefont {Ando}}, \ and\ \bibinfo {author} {\bibfnamefont
  {T.}~\bibnamefont {Sato}},\ }\href {\doibase 10.1103/PhysRevB.99.245105}
  {\bibfield  {journal} {\bibinfo  {journal} {Phys. Rev. B}\ }\textbf {\bibinfo
  {volume} {99}},\ \bibinfo {pages} {245105} (\bibinfo {year}
  {2019})}\BibitemShut {NoStop}%
\bibitem [{\citenamefont {Hu}\ \emph {et~al.}(2016)\citenamefont {Hu},
  \citenamefont {Tang}, \citenamefont {Liu}, \citenamefont {Liu}, \citenamefont
  {Zhu}, \citenamefont {Graf}, \citenamefont {Myhro}, \citenamefont {Tran},
  \citenamefont {Lau}, \citenamefont {Wei},\ and\ \citenamefont
  {Mao}}]{DNL-ZrSiSeandZrSiTe}%
  \BibitemOpen
  \bibfield  {author} {\bibinfo {author} {\bibfnamefont {J.}~\bibnamefont
  {Hu}}, \bibinfo {author} {\bibfnamefont {Z.}~\bibnamefont {Tang}}, \bibinfo
  {author} {\bibfnamefont {J.}~\bibnamefont {Liu}}, \bibinfo {author}
  {\bibfnamefont {X.}~\bibnamefont {Liu}}, \bibinfo {author} {\bibfnamefont
  {Y.}~\bibnamefont {Zhu}}, \bibinfo {author} {\bibfnamefont {D.}~\bibnamefont
  {Graf}}, \bibinfo {author} {\bibfnamefont {K.}~\bibnamefont {Myhro}},
  \bibinfo {author} {\bibfnamefont {S.}~\bibnamefont {Tran}}, \bibinfo {author}
  {\bibfnamefont {C.~N.}\ \bibnamefont {Lau}}, \bibinfo {author} {\bibfnamefont
  {J.}~\bibnamefont {Wei}}, \ and\ \bibinfo {author} {\bibfnamefont
  {Z.}~\bibnamefont {Mao}},\ }\href {\doibase 10.1103/PhysRevLett.117.016602}
  {\bibfield  {journal} {\bibinfo  {journal} {Phys. Rev. Lett.}\ }\textbf
  {\bibinfo {volume} {117}},\ \bibinfo {pages} {016602} (\bibinfo {year}
  {2016})}\BibitemShut {NoStop}%
\bibitem [{\citenamefont {Fu}\ \emph {et~al.}(2019)\citenamefont {Fu},
  \citenamefont {Yi}, \citenamefont {Zhang}, \citenamefont {Caputo},
  \citenamefont {Ma}, \citenamefont {Gao}, \citenamefont {Lv}, \citenamefont
  {Kong}, \citenamefont {Huang}, \citenamefont {Richard}, \citenamefont {Shi},
  \citenamefont {Strocov}, \citenamefont {Fang}, \citenamefont {Weng},
  \citenamefont {Shi}, \citenamefont {Qian},\ and\ \citenamefont
  {Ding}}]{DNL-ZrSiS}%
  \BibitemOpen
  \bibfield  {author} {\bibinfo {author} {\bibfnamefont {B.-B.}\ \bibnamefont
  {Fu}}, \bibinfo {author} {\bibfnamefont {C.-J.}\ \bibnamefont {Yi}}, \bibinfo
  {author} {\bibfnamefont {T.-T.}\ \bibnamefont {Zhang}}, \bibinfo {author}
  {\bibfnamefont {M.}~\bibnamefont {Caputo}}, \bibinfo {author} {\bibfnamefont
  {J.-Z.}\ \bibnamefont {Ma}}, \bibinfo {author} {\bibfnamefont
  {X.}~\bibnamefont {Gao}}, \bibinfo {author} {\bibfnamefont {B.~Q.}\
  \bibnamefont {Lv}}, \bibinfo {author} {\bibfnamefont {L.-Y.}\ \bibnamefont
  {Kong}}, \bibinfo {author} {\bibfnamefont {Y.-B.}\ \bibnamefont {Huang}},
  \bibinfo {author} {\bibfnamefont {P.}~\bibnamefont {Richard}}, \bibinfo
  {author} {\bibfnamefont {M.}~\bibnamefont {Shi}}, \bibinfo {author}
  {\bibfnamefont {V.~N.}\ \bibnamefont {Strocov}}, \bibinfo {author}
  {\bibfnamefont {C.}~\bibnamefont {Fang}}, \bibinfo {author} {\bibfnamefont
  {H.-M.}\ \bibnamefont {Weng}}, \bibinfo {author} {\bibfnamefont {Y.-G.}\
  \bibnamefont {Shi}}, \bibinfo {author} {\bibfnamefont {T.}~\bibnamefont
  {Qian}}, \ and\ \bibinfo {author} {\bibfnamefont {H.}~\bibnamefont {Ding}},\
  }\href {\doibase 10.1126/sciadv.aau6459} {\bibfield  {journal} {\bibinfo
  {journal} {Science Advances}\ }\textbf {\bibinfo {volume} {5}},\ \bibinfo
  {pages} {eaau6459} (\bibinfo {year} {2019})}\BibitemShut {NoStop}%
\bibitem [{\citenamefont {Lou}\ \emph {et~al.}(2016)\citenamefont {Lou},
  \citenamefont {Ma}, \citenamefont {Xu}, \citenamefont {Fu}, \citenamefont
  {Kong}, \citenamefont {Shi}, \citenamefont {Richard}, \citenamefont {Weng},
  \citenamefont {Fang}, \citenamefont {Sun}, \citenamefont {Wang},
  \citenamefont {Lei}, \citenamefont {Qian}, \citenamefont {Ding},\ and\
  \citenamefont {Wang}}]{TI-ZrSnTe}%
  \BibitemOpen
  \bibfield  {author} {\bibinfo {author} {\bibfnamefont {R.}~\bibnamefont
  {Lou}}, \bibinfo {author} {\bibfnamefont {J.-Z.}\ \bibnamefont {Ma}},
  \bibinfo {author} {\bibfnamefont {Q.-N.}\ \bibnamefont {Xu}}, \bibinfo
  {author} {\bibfnamefont {B.-B.}\ \bibnamefont {Fu}}, \bibinfo {author}
  {\bibfnamefont {L.-Y.}\ \bibnamefont {Kong}}, \bibinfo {author}
  {\bibfnamefont {Y.-G.}\ \bibnamefont {Shi}}, \bibinfo {author} {\bibfnamefont
  {P.}~\bibnamefont {Richard}}, \bibinfo {author} {\bibfnamefont {H.-M.}\
  \bibnamefont {Weng}}, \bibinfo {author} {\bibfnamefont {Z.}~\bibnamefont
  {Fang}}, \bibinfo {author} {\bibfnamefont {S.-S.}\ \bibnamefont {Sun}},
  \bibinfo {author} {\bibfnamefont {Q.}~\bibnamefont {Wang}}, \bibinfo {author}
  {\bibfnamefont {H.-C.}\ \bibnamefont {Lei}}, \bibinfo {author} {\bibfnamefont
  {T.}~\bibnamefont {Qian}}, \bibinfo {author} {\bibfnamefont {H.}~\bibnamefont
  {Ding}}, \ and\ \bibinfo {author} {\bibfnamefont {S.-C.}\ \bibnamefont
  {Wang}},\ }\href {\doibase 10.1103/PhysRevB.93.241104} {\bibfield  {journal}
  {\bibinfo  {journal} {Phys. Rev. B}\ }\textbf {\bibinfo {volume} {93}},\
  \bibinfo {pages} {241104(R)} (\bibinfo {year} {2016})}\BibitemShut {NoStop}%
\bibitem [{\citenamefont {Singha}\ \emph {et~al.}(2017)\citenamefont {Singha},
  \citenamefont {Pariari}, \citenamefont {Satpati},\ and\ \citenamefont
  {Mandal}}]{Singha2017}%
  \BibitemOpen
  \bibfield  {author} {\bibinfo {author} {\bibfnamefont {R.}~\bibnamefont
  {Singha}}, \bibinfo {author} {\bibfnamefont {A.}~\bibnamefont {Pariari}},
  \bibinfo {author} {\bibfnamefont {B.}~\bibnamefont {Satpati}}, \ and\
  \bibinfo {author} {\bibfnamefont {P.}~\bibnamefont {Mandal}},\ }\href
  {\doibase 10.1103/PhysRevB.96.245138} {\bibfield  {journal} {\bibinfo
  {journal} {Phys. Rev. B}\ }\textbf {\bibinfo {volume} {96}},\ \bibinfo
  {pages} {245138} (\bibinfo {year} {2017})}\BibitemShut {NoStop}%
\bibitem [{\citenamefont {Lv}\ \emph {et~al.}(2019)\citenamefont {Lv},
  \citenamefont {Chen}, \citenamefont {Qiao}, \citenamefont {Ma},\ and\
  \citenamefont {Yang}}]{Lv2019}%
  \BibitemOpen
  \bibfield  {author} {\bibinfo {author} {\bibfnamefont {B.}~\bibnamefont
  {Lv}}, \bibinfo {author} {\bibfnamefont {J.}~\bibnamefont {Chen}}, \bibinfo
  {author} {\bibfnamefont {L.}~\bibnamefont {Qiao}}, \bibinfo {author}
  {\bibfnamefont {J.}~\bibnamefont {Ma}}, \ and\ \bibinfo {author}
  {\bibfnamefont {X.}~\bibnamefont {Yang}},\ }\href
  {https://doi.org/10.1088/1361-648X/ab2498} {\bibfield  {journal} {\bibinfo
  {journal} {J.phys.:condens.matter}\ }\textbf {\bibinfo {volume} {31}},\
  \bibinfo {pages} {355601} (\bibinfo {year} {2019})}\BibitemShut {NoStop}%
\bibitem [{\citenamefont {Hosen}\ \emph {et~al.}(2018)\citenamefont {Hosen},
  \citenamefont {Dhakal}, \citenamefont {Dimitri}, \citenamefont {Maldonado},
  \citenamefont {Aperis}, \citenamefont {Kabir}, \citenamefont {Sims},
  \citenamefont {Riseborough}, \citenamefont {Oppeneer}, \citenamefont
  {Kaczorowski}, \citenamefont {Durakiewicz},\ and\ \citenamefont
  {Neupane}}]{Hosen2018}%
  \BibitemOpen
  \bibfield  {author} {\bibinfo {author} {\bibfnamefont {M.~M.}\ \bibnamefont
  {Hosen}}, \bibinfo {author} {\bibfnamefont {G.}~\bibnamefont {Dhakal}},
  \bibinfo {author} {\bibfnamefont {K.}~\bibnamefont {Dimitri}}, \bibinfo
  {author} {\bibfnamefont {P.}~\bibnamefont {Maldonado}}, \bibinfo {author}
  {\bibfnamefont {A.}~\bibnamefont {Aperis}}, \bibinfo {author} {\bibfnamefont
  {F.}~\bibnamefont {Kabir}}, \bibinfo {author} {\bibfnamefont
  {C.}~\bibnamefont {Sims}}, \bibinfo {author} {\bibfnamefont {P.}~\bibnamefont
  {Riseborough}}, \bibinfo {author} {\bibfnamefont {P.~M.}\ \bibnamefont
  {Oppeneer}}, \bibinfo {author} {\bibfnamefont {D.}~\bibnamefont
  {Kaczorowski}}, \bibinfo {author} {\bibfnamefont {T.}~\bibnamefont
  {Durakiewicz}}, \ and\ \bibinfo {author} {\bibfnamefont {M.}~\bibnamefont
  {Neupane}},\ }\href {\doibase 10.1038/s41598-018-31296-7} {\bibfield
  {journal} {\bibinfo  {journal} {Scientific Reports}\ }\textbf {\bibinfo
  {volume} {8}},\ \bibinfo {pages} {13283} (\bibinfo {year}
  {2018})}\BibitemShut {NoStop}%
\bibitem [{\citenamefont {Yang}\ \emph {et~al.}(2020)\citenamefont {Yang},
  \citenamefont {Qian}, \citenamefont {Yan}, \citenamefont {Li}, \citenamefont
  {Song}, \citenamefont {Wang}, \citenamefont {Yi}, \citenamefont {Feng},
  \citenamefont {Weng},\ and\ \citenamefont {Shi}}]{Yang2020}%
  \BibitemOpen
  \bibfield  {author} {\bibinfo {author} {\bibfnamefont {M.}~\bibnamefont
  {Yang}}, \bibinfo {author} {\bibfnamefont {Y.}~\bibnamefont {Qian}}, \bibinfo
  {author} {\bibfnamefont {D.}~\bibnamefont {Yan}}, \bibinfo {author}
  {\bibfnamefont {Y.}~\bibnamefont {Li}}, \bibinfo {author} {\bibfnamefont
  {Y.}~\bibnamefont {Song}}, \bibinfo {author} {\bibfnamefont {Z.}~\bibnamefont
  {Wang}}, \bibinfo {author} {\bibfnamefont {C.}~\bibnamefont {Yi}}, \bibinfo
  {author} {\bibfnamefont {H.~L.}\ \bibnamefont {Feng}}, \bibinfo {author}
  {\bibfnamefont {H.}~\bibnamefont {Weng}}, \ and\ \bibinfo {author}
  {\bibfnamefont {Y.}~\bibnamefont {Shi}},\ }\href {\doibase
  10.1103/PhysRevMaterials.4.094203} {\bibfield  {journal} {\bibinfo  {journal}
  {Physical Review Materials}\ }\textbf {\bibinfo {volume} {4}},\ \bibinfo
  {pages} {094203} (\bibinfo {year} {2020})}\BibitemShut {NoStop}%
\bibitem [{\citenamefont {Yue}\ \emph {et~al.}(2020)\citenamefont {Yue},
  \citenamefont {Qian}, \citenamefont {Yang}, \citenamefont {Geng},
  \citenamefont {Yi}, \citenamefont {Kumar}, \citenamefont {Shimada},
  \citenamefont {Cheng}, \citenamefont {Chen}, \citenamefont {Wang},
  \citenamefont {Weng}, \citenamefont {Shi}, \citenamefont {Wu},\ and\
  \citenamefont {Feng}}]{Yue2020}%
  \BibitemOpen
  \bibfield  {author} {\bibinfo {author} {\bibfnamefont {S.}~\bibnamefont
  {Yue}}, \bibinfo {author} {\bibfnamefont {Y.}~\bibnamefont {Qian}}, \bibinfo
  {author} {\bibfnamefont {M.}~\bibnamefont {Yang}}, \bibinfo {author}
  {\bibfnamefont {D.}~\bibnamefont {Geng}}, \bibinfo {author} {\bibfnamefont
  {C.}~\bibnamefont {Yi}}, \bibinfo {author} {\bibfnamefont {S.}~\bibnamefont
  {Kumar}}, \bibinfo {author} {\bibfnamefont {K.}~\bibnamefont {Shimada}},
  \bibinfo {author} {\bibfnamefont {P.}~\bibnamefont {Cheng}}, \bibinfo
  {author} {\bibfnamefont {L.}~\bibnamefont {Chen}}, \bibinfo {author}
  {\bibfnamefont {Z.}~\bibnamefont {Wang}}, \bibinfo {author} {\bibfnamefont
  {H.}~\bibnamefont {Weng}}, \bibinfo {author} {\bibfnamefont {Y.}~\bibnamefont
  {Shi}}, \bibinfo {author} {\bibfnamefont {K.}~\bibnamefont {Wu}}, \ and\
  \bibinfo {author} {\bibfnamefont {B.}~\bibnamefont {Feng}},\ }\href {\doibase
  10.1103/PhysRevB.102.155109} {\bibfield  {journal} {\bibinfo  {journal}
  {Physical Review B}\ }\textbf {\bibinfo {volume} {102}},\ \bibinfo {pages}
  {155109} (\bibinfo {year} {2020})}\BibitemShut {NoStop}%
\bibitem [{\citenamefont {Petrovic}\ \emph {et~al.}(2001)\citenamefont
  {Petrovic}, \citenamefont {Pagliuso}, \citenamefont {Hundley}, \citenamefont
  {Movshovich}, \citenamefont {Sarrao}, \citenamefont {Thompson}, \citenamefont
  {Fisk},\ and\ \citenamefont {Monthoux}}]{Petrovic_2001}%
  \BibitemOpen
  \bibfield  {author} {\bibinfo {author} {\bibfnamefont {C.}~\bibnamefont
  {Petrovic}}, \bibinfo {author} {\bibfnamefont {P.~G.}\ \bibnamefont
  {Pagliuso}}, \bibinfo {author} {\bibfnamefont {M.~F.}\ \bibnamefont
  {Hundley}}, \bibinfo {author} {\bibfnamefont {R.}~\bibnamefont {Movshovich}},
  \bibinfo {author} {\bibfnamefont {J.~L.}\ \bibnamefont {Sarrao}}, \bibinfo
  {author} {\bibfnamefont {J.~D.}\ \bibnamefont {Thompson}}, \bibinfo {author}
  {\bibfnamefont {Z.}~\bibnamefont {Fisk}}, \ and\ \bibinfo {author}
  {\bibfnamefont {P.}~\bibnamefont {Monthoux}},\ }\href {\doibase
  10.1088/0953-8984/13/17/103} {\bibfield  {journal} {\bibinfo  {journal}
  {Journal of Physics: Condensed Matter}\ }\textbf {\bibinfo {volume} {13}},\
  \bibinfo {pages} {L337} (\bibinfo {year} {2001})}\BibitemShut {NoStop}%
\bibitem [{\citenamefont {Li}\ \emph {et~al.}(2021)\citenamefont {Li},
  \citenamefont {Lv}, \citenamefont {Fang}, \citenamefont {Guo}, \citenamefont
  {Wu}, \citenamefont {Wu},\ and\ \citenamefont {Shen}}]{Li2021}%
  \BibitemOpen
  \bibfield  {author} {\bibinfo {author} {\bibfnamefont {P.}~\bibnamefont
  {Li}}, \bibinfo {author} {\bibfnamefont {B.}~\bibnamefont {Lv}}, \bibinfo
  {author} {\bibfnamefont {Y.}~\bibnamefont {Fang}}, \bibinfo {author}
  {\bibfnamefont {W.}~\bibnamefont {Guo}}, \bibinfo {author} {\bibfnamefont
  {Z.}~\bibnamefont {Wu}}, \bibinfo {author} {\bibfnamefont {Y.}~\bibnamefont
  {Wu}}, \ and\ \bibinfo {author} {\bibfnamefont {D.}~\bibnamefont {Shen}},\
  }\href {\doibase https://doi.org/10.1007/s11433-020-1642-2} {\bibfield
  {journal} {\bibinfo  {journal} {Science China Physics, Mechanics {\&}
  Astronomy}\ }\textbf {\bibinfo {volume} {64}},\ \bibinfo {pages} {237412}
  (\bibinfo {year} {2021})}\BibitemShut {NoStop}%
\bibitem [{\citenamefont {Blaha}\ \emph {et~al.}()\citenamefont {Blaha},
  \citenamefont {Schwarz}, \citenamefont {Madsen}, \citenamefont {Kvasnicka},\
  and\ \citenamefont {Luitz}}]{S3}%
  \BibitemOpen
  \bibfield  {author} {\bibinfo {author} {\bibfnamefont {P.}~\bibnamefont
  {Blaha}}, \bibinfo {author} {\bibfnamefont {K.}~\bibnamefont {Schwarz}},
  \bibinfo {author} {\bibfnamefont {G.~K.}\ \bibnamefont {Madsen}}, \bibinfo
  {author} {\bibfnamefont {D.}~\bibnamefont {Kvasnicka}}, \ and\ \bibinfo
  {author} {\bibfnamefont {J.}~\bibnamefont {Luitz}},\ }\href@noop {} {\bibinfo
   {journal} {WIEN2K, An Augmented Plane Wave+ Local Orbitals Program for
  Calculating Crystal Properties, (Karlheinz Schwarz, Vienna University of
  Technology, Austria, 2001)}\ }\BibitemShut {NoStop}%
\bibitem [{\citenamefont {Perdew}\ \emph {et~al.}(1996)\citenamefont {Perdew},
  \citenamefont {Burke},\ and\ \citenamefont
  {Ernzerhof}}]{PhysRevLett.77.3865}%
  \BibitemOpen
\bibfield  {journal} {  }\bibfield  {author} {\bibinfo {author} {\bibfnamefont
  {J.~P.}\ \bibnamefont {Perdew}}, \bibinfo {author} {\bibfnamefont
  {K.}~\bibnamefont {Burke}}, \ and\ \bibinfo {author} {\bibfnamefont
  {M.}~\bibnamefont {Ernzerhof}},\ }\href {\doibase
  10.1103/PhysRevLett.77.3865} {\bibfield  {journal} {\bibinfo  {journal}
  {Phys. Rev. Lett.}\ }\textbf {\bibinfo {volume} {77}},\ \bibinfo {pages}
  {3865} (\bibinfo {year} {1996})}\BibitemShut {NoStop}%
\bibitem [{\citenamefont {Hu}\ \emph {et~al.}(2011)\citenamefont {Hu},
  \citenamefont {Zheng}, \citenamefont {Yuan}, \citenamefont {Dong},
  \citenamefont {Cheng}, \citenamefont {Chen},\ and\ \citenamefont
  {Wang}}]{Hu2011a}%
  \BibitemOpen
  \bibfield  {author} {\bibinfo {author} {\bibfnamefont {B.~F.}\ \bibnamefont
  {Hu}}, \bibinfo {author} {\bibfnamefont {P.}~\bibnamefont {Zheng}}, \bibinfo
  {author} {\bibfnamefont {R.~H.}\ \bibnamefont {Yuan}}, \bibinfo {author}
  {\bibfnamefont {T.}~\bibnamefont {Dong}}, \bibinfo {author} {\bibfnamefont
  {B.}~\bibnamefont {Cheng}}, \bibinfo {author} {\bibfnamefont {Z.~G.}\
  \bibnamefont {Chen}}, \ and\ \bibinfo {author} {\bibfnamefont {N.~L.}\
  \bibnamefont {Wang}},\ }\href {\doibase 10.1103/PhysRevB.83.155113}
  {\bibfield  {journal} {\bibinfo  {journal} {Physical Review B}\ }\textbf
  {\bibinfo {volume} {83}},\ \bibinfo {pages} {155113} (\bibinfo {year}
  {2011})}\BibitemShut {NoStop}%
\bibitem [{\citenamefont {Chen}\ \emph {et~al.}(2015)\citenamefont {Chen},
  \citenamefont {Zhang}, \citenamefont {Schneeloch}, \citenamefont {Zhang},
  \citenamefont {Li}, \citenamefont {Gu},\ and\ \citenamefont
  {Wang}}]{Chen2015}%
  \BibitemOpen
  \bibfield  {author} {\bibinfo {author} {\bibfnamefont {R.~Y.}\ \bibnamefont
  {Chen}}, \bibinfo {author} {\bibfnamefont {S.~J.}\ \bibnamefont {Zhang}},
  \bibinfo {author} {\bibfnamefont {J.~A.}\ \bibnamefont {Schneeloch}},
  \bibinfo {author} {\bibfnamefont {C.}~\bibnamefont {Zhang}}, \bibinfo
  {author} {\bibfnamefont {Q.}~\bibnamefont {Li}}, \bibinfo {author}
  {\bibfnamefont {G.~D.}\ \bibnamefont {Gu}}, \ and\ \bibinfo {author}
  {\bibfnamefont {N.~L.}\ \bibnamefont {Wang}},\ }\href {\doibase
  10.1103/PhysRevB.92.075107} {\bibfield  {journal} {\bibinfo  {journal}
  {Physical Review B}\ }\textbf {\bibinfo {volume} {92}},\ \bibinfo {pages}
  {075107} (\bibinfo {year} {2015})}\BibitemShut {NoStop}%
\bibitem [{\citenamefont {Pfuner}\ \emph {et~al.}(2010)\citenamefont {Pfuner},
  \citenamefont {Lerch}, \citenamefont {Chu}, \citenamefont {Kuo},
  \citenamefont {Fisher},\ and\ \citenamefont {Degiorgi}}]{HoTe3}%
  \BibitemOpen
  \bibfield  {author} {\bibinfo {author} {\bibfnamefont {F.}~\bibnamefont
  {Pfuner}}, \bibinfo {author} {\bibfnamefont {P.}~\bibnamefont {Lerch}},
  \bibinfo {author} {\bibfnamefont {J.-H.}\ \bibnamefont {Chu}}, \bibinfo
  {author} {\bibfnamefont {H.-H.}\ \bibnamefont {Kuo}}, \bibinfo {author}
  {\bibfnamefont {I.~R.}\ \bibnamefont {Fisher}}, \ and\ \bibinfo {author}
  {\bibfnamefont {L.}~\bibnamefont {Degiorgi}},\ }\href {\doibase
  10.1103/PhysRevB.81.195110} {\bibfield  {journal} {\bibinfo  {journal} {Phys.
  Rev. B}\ }\textbf {\bibinfo {volume} {81}},\ \bibinfo {pages} {195110}
  (\bibinfo {year} {2010})}\BibitemShut {NoStop}%
\end{thebibliography}%


%

\end{document}